%% file: bulgeKinematics2018.tex
\documentclass[fleqn,usenatbib]{mnras}

\usepackage{newtxtext,newtxmath}

\usepackage[T1]{fontenc}
\usepackage{ae,aecompl}

\usepackage{graphicx}	
\usepackage{amsmath}	
\usepackage{amssymb}	

\usepackage{subfigure}
\usepackage{placeins}
\usepackage{courier}
\usepackage{verbatim}
\usepackage{paralist}

\usepackage{ulem}
\usepackage{float}
\usepackage{arydshln}

\title[Milky Way barred bulge kinematics.]{The Milky Way bar/bulge in proper motions: a 3D view from VIRAC \& \textit{Gaia}}

\author[J. P. Clarke et al.]{
Jonathan P. Clarke,$^{1}$\thanks{E-mail: jclarke@mpe.mpg.de (JPC)}
Christopher Wegg,$^{1,2}$
Ortwin Gerhard,$^{1}$\thanks{E-mail: gerhard@mpe.mpg.de (OG)}
Leigh C. Smith,$^{3,4}$ \newauthor
Phil W. Lucas$^{4}$
and Shola M. Wylie,$^{1}$
\\
$^{1}$ Max-Planck-Institut f{\"u}r Extraterrestrische Physik, Gie{\ss}enbachstra{\ss}e, D-85748 Garching, Germany \\
$^{2}$ Laboratoire Lagrange, Universit\'e C\^ote d' Azur, Observatoire de la C\^ote d' Azur, CNRS, Bd de l'Observatoire, 06304 Nice, France \\
$^{3}$ Institute of Astronomy, University of Cambridge, Madingley Rd, Cambridge, CB3 0HA, UK\\
$^{4}$ School of Physics, Astronomy and Mathematics, University of Hertfordshire, College Lane, Hatfield AL10 9AB, UK
}

\date{Accepted XXX. Received YYY; in original form ZZZ}

\pubyear{2019}

\newcommand{\Lagr}{\mathcal{L}}

\newcommand{\mul}{\mu_l^\star}
\newcommand{\mub}{\mu_b}
\newcommand{\mpml}{<\mu_l^\star>}
\newcommand{\dmpml}{\Delta<\mu_l^\star>}
\newcommand{\mpmb}{<\mu_b>}
\newcommand{\dpml}{\sigma_{\mu_l^\star}}
\newcommand{\dpmb}{\sigma_{\mu_b}}

\newcommand{\dg}{^\circ}
\newcommand{\anisotropy}{ \dpml / \dpmb }

\newcommand{\kms}{ \mathrm{ km \, s^{-1}} }
\newcommand{\kmskpc}{ \mathrm{ \kms \, kpc^{-1} } }
\newcommand{\masyr}{\mathrm{mas \, yr^{-1}}}

\newcommand{\mk}{M_{K_{s0}}}

\begin{document}
\label{firstpage}
\pagerange{\pageref{firstpage}--\pageref{lastpage}}
\maketitle

\defcitealias{portail_2017}{P17}
\defcitealias{smith_2018}{S18}
\defcitealias{wegg_2013}{W13}
\defcitealias{wegg_2015}{W15}
\defcitealias{rattenbury_2007}{R07}
\defcitealias{kozlowski_2006}{K06}

\begin{abstract}
We have derived absolute proper motions of the entire Galactic bulge region from VIRAC and \textit{Gaia}. We present these as both integrated on-sky maps and, after isolating standard candle red clump (RC) stars, as a function of distance using RC magnitude as a proxy.
These data provide a new global, 3-dimensional view of the Milky Way barred bulge kinematics.
We find a gradient in the mean longitudinal proper motion, $\mpml$, between the different sides of the bar, which is sensitive to the bar pattern speed. The split RC has distinct proper motions and is colder than other stars at similar distance. The proper motion correlation map has a quadrupole pattern in all magnitude slices showing no evidence for a separate, more axisymmetric inner bulge component.
The line-of-sight integrated kinematic maps show a high central velocity dispersion surrounded by a more asymmetric dispersion profile. $ \sigma_{\mu_l} / \sigma_{\mu_b} $ is smallest, $\sim1.1$, near the minor axis and reaches $\sim1.4$ near the disc plane. The integrated $\mpmb$ pattern signals a superposition of bar rotation and internal streaming motion, with the near part shrinking in latitude and the far part expanding.
To understand and interpret these remarkable data, we compare to a made-to-measure barred dynamical model, folding in the VIRAC selection function to construct mock maps. 
We find that our model of the barred bulge, with a pattern speed of 37.5 $\kmskpc$, is able to reproduce all observed features impressively well. 
Dynamical models like this will be key to unlocking the full potential of these data.

\end{abstract}





\section{Introduction}\label{sec:intro}
\input{sections/a_KinematicsIntro.tex}

\section{VVV proper motions}\label{sec:vvvpm}
\input{sections/b_KinematicsVVVpm.tex}

\section{Made-to-Measure Milky Way Models}\label{sec:m2m}

\input{sections/c_KinematicsNmagic.tex}


\section{Red Giant Kinematics}\label{sec:rg_kinematics}

\input{sections/d_RGB_kinematics.tex}

\input{sections/e_correlationRG.tex}


\section{Extracting the RC\&B from the VIRAC RGB}\label{sec:getRCB}

\input{sections/f_KinematicsGetRc.tex}


\section{Red Clump Kinematics }\label{sec:rc_kinematics}

\input{sections/g_kinematics_Bsliced.tex}

\input{sections/h_kinematics_Ksliced.tex}

\section{Summary \& Conclusions}\label{sec:conclusions}
\input{sections/z_KinematicsConclusion.tex}

\section*{Acknowledgements}
We acknowledge the simultaneous work by \citet{sanders_2019}, who also used an absolute proper motion catalogue derived from VVV and \textit{Gaia} DR2 to study the kinematics of the bulge. The authors of both publications were aware of each others work, but arrived at their conclusions independently.
We thank the anonymous referee whose comments led to improvements in the paper.
We gratefully acknowledge the pioneering work of Matthieu Portail in producing the current version and documentation of \texttt{NMAGIC} used in this publication.
We acknowledge useful discussions with Isabella S{\"o}ldner-Rembold and Johanna Hartke.
CW acknowledges funding from the European Union's Horizon 2020 research and innovation program under the Marie Sk\l{}odowska-Curie grant agreement No 798384.
This work was based on data products from observations made with ESO Telescopes at the La Silla or Paranal Observatories under ESO programme ID 179.B-2002.
We are grateful to the VISTA Science Archive for providing a user friendly interface from which we could access the VIRAC catalogue.
This work presents results from the European Space Agency (ESA) space mission Gaia. Gaia data are being processed by the Gaia Data Processing and Analysis Consortium (DPAC). Funding for the DPAC is provided by national institutions, in particular the institutions participating in the Gaia MultiLateral Agreement (MLA). The Gaia mission website is https://www.cosmos.esa.int/gaia. The Gaia archive website is https://archives.esac.esa.int/gaia.
We have used the python $astropy.coordinates.SkyCoord$ package to convert coordinates and proper motions between coordinate systems and the $cov\_pmrapmdec\_to\_pmllpmbb$ function from $galpy$ \citep{bovy_2015} to convert the error covariance matrix between coordinate systems.



\bibliographystyle{mnras}
\bibliography{bulgeKinematics2018} 



\appendix


\bsp	
\label{lastpage}

\end{document}

%% file: sections/a_KinematicsIntro.tex

The Milky Way (MW) is a barred galaxy with a boxy/peanut bulge, which appears to be in a relatively late stage of evolution based on its low specific star formation rate \citep[see][]{bland_hawthorn_2016}. The presence of the bar was first convincingly shown in the 1990s through its effect on the distribution and kinematics of stars and gas \citep{binney_1991,stanek_1994,weiland_1994,zhao_1994,fux_1999}. It is now well established that a dominant fraction of the MW bulge is composed of a triaxial bar structure \citep{lopez_corredoira_2005,rattenbury_2007b,saito_2011,wegg_2013}. 
There is still an ongoing debate as to whether there exists a secondary classical bulge component in the central parts of the bulge \citep{shen_2010,rojas_arriagada_2017,di_matteo_2015,barbuy_2018}. With modern stellar surveys, the MW bulge and bar can be studied at great depth, rapidly making the MW a prototypical system for understanding the
formation and evolution of similar galaxies.

A prominent feature of the barred bulge is the split red clump (RC) which was first reported by \citet{nataf_2010,mcwilliam_2010} using OGLE-III photometry and 2MASS data respectively. 
They showed that this phenomenon occurs close to the MW minor axis at latitudes of $|b|\gtrsim5\dg$. 
From these analyses it was suggested that the split RC could be the result of a funnel shaped component in the bulge which is now commonly referred to as X-shaped.
Further evidence for this scenario was presented by
\begin{inparaenum}
    \item \citet{saito_2011} also using 2MASS data who observed the X-shape within $|l|<2\dg$ with the two density peaks merging at latitudes $|b|<4^\circ$;
    \item \citet{ness_2012} who showed that 2 ARGOS fields for which $b<-5\dg$ exhibit this bi-modal magnitude distribution only for stars with [Fe/H] $>$ 0.5;
    \item \citet[hereafter W13]{wegg_2013} who reconstructed the full 3D density of RC stars using star counts from the VVV survey;
    \item \citet{nataf_2015} who compared OGLE-III photometry to two barred N-body models that both show the split RC at high latitudes;
    \item \citet{ness_2016} who used WISE images to demonstrate the X-shape morphology of the MW bulge in projection; and
    \item \citet{gonzalez_2016} who compared the X-shape bulge of NGC 4710 from MUSE with that of the MW and found general agreement.
\end{inparaenum}
Such peanut shaped bulges have been observed in external galaxies \citep{lutticke_2000,bureau_2006,laurikainen_2014} and naturally form in N-body simulations due to the buckling instability and/or orbits in vertical resonance \citep{combes_1990,raha_1991,athanassoula_2005,debattista_2006}.
An alternative explanation for the split RC was proposed by \citet{lee_YW_2015,lee_YW_2018} who suggested that the split RC we observe is not due to a bi-modal density profile but rather that it is due to a population effect. Their model contains a bar superimposed on top of a classical bulge with two RC populations.
The RC is so prominent in the literature because its narrow range of absolute magnitudes makes their apparent magnitude a good proxy for distance \citep{stanek_1994}.

There have been many previous proper motion studies in the galactic bulge  (\citealt{spaenhauer_1992,kozlowski_2006,rattenbury_2007,soto_2014,clarkson_2018} and references therein). 
This work has highlighted gradients in the proper motion dispersions, $\dpml$ and $\dpmb$, see in particular \citet[hereafter K06]{kozlowski_2006} and \citet[hereafter R07]{rattenbury_2007}, and measured the proper motion dispersion ratio, $\anisotropy$, $\sim 1.2$ in near galactic center fields in the (+$l$,-$b$) quadrant, see also figure \ref{fig:compare2otherData} below. Due to a lack of background quasars to anchor the proper motion reference frame, these studies had to work with relative proper motions. Moreover, the relatively low numbers of stars in these studies restricted them to investigating only projected kinematics.

Recent and ongoing large scale surveys such as OGLE, UKIDSS, 2MASS, VVV, ARGOS, BRAVA, GES, GIBS and APOGEE allow bulge studies to extend beyond integrated LOS measurements and probe the bulge as a function of distance.
Using VISTA Variables in the Via Lactea (VVV) DR1 \citep{saito_2012} star counts \citetalias{wegg_2013} performed a 3D density mapping of the galactic bulge. They found a strongly boxy/peanut shaped bulge, with a prominent X-shape, and the major axis of the bar tilted by $(27\pm2)\dg$ to the line of sight. \citet[hereafter W15]{wegg_2015} followed this up studying the long bar that extends beyond the MW bulge and concluded that the central boxy/peanut bulge is the more vertically extended counterpart to the long bar. This suggests that the two structures are dynamically related and share a common origin although this requires further confirmation.
It has also been possible to study the MW bulge in 3D with radial velocities. \citet{vasquez_2013} observed a sample of 454 bulge giants in a region at $(l=0.\dg,b=-6.\dg)$ with stars well distributed over the bright and faint RC peaks. They found evidence of streaming motions within the bar with an excess of stars in the bright RC moving towards the sun and the converse for the faint RC. This streaming motion is in the same sense as the bar rotates.

The VVV Infrared Astrometric Catalogue (VIRAC) \citep[hereafter S18]{smith_2018} has provided a total of $\sim175\,000\,000$ proper motion measurements across the Galactic bulge region, ($-10\, <\,l/\mathrm{deg}\,<\,10$, $-10\,<\,b/\mathrm{deg}\,<\,5$). 
Combined with data from \textit{Gaia} \citep{brown_2018} to provide an absolute reference frame, these data offer an unprecedented opportunity to study the 3D proper motion structure of the MW bulge.
The goal of this paper is to derive LOS integrated and distance-resolved maps of mean proper motions and dispersions from the VIRAC data and use a dynamical model to aid in their interpretation.

Dynamical models are a key tool in interpreting the vast quantity of data now being provided by large stellar surveys. \citet[hereafter P17]{portail_2017} used the made-to-measure (M2M) method to construct barred dynamical models fit to VVV, UKIDSS, 2MASS, BRAVA and ARGOS. These models have well defined pattern speeds and \citetalias{portail_2017} found the best fitting pattern speed to be $\Omega=39.0\pm3.5 $ $\kmskpc$. They also found dynamical evidence for a centrally concentrated nuclear disc of mass $\sim0.2 \times 10^{10}$ $M_\odot$. 
This extra mass is required to better match the inner BRAVA dispersions and the OGLE $b$ proper motions presented by \citetalias{rattenbury_2007}. Additionally the best fitting models favour a core/shallow cusp in the dark matter within the bulge region. These models are in good agreement with all the data to which they were fitted, making them a specialised tool for studying the MW bulge. We use them here to predict proper motion kinematics.

The paper is organised as follows. 
In section \ref{sec:vvvpm} we extract a colour selected sample of red giant branch (RGB) stars with absolute proper motions from VIRAC and \textit{Gaia}. 
Section \ref{sec:m2m} describes the modelling approach to observe the \citetalias{portail_2017} M2M  model in a manner consistent with our VIRAC subsample. 
In section \ref{sec:rg_kinematics} we present integrated on-sky maps of the mean proper motions, proper motion dispersions, dispersion ratio and proper motion correlation. 
Section \ref{sec:getRCB} discusses the method to extract a statistical sample of RC stars together with the red giant branch bump (RGBB) and asymptotic giant branch bump (AGBB) stars for use as a distance proxy. 
In section \ref{sec:rc_kinematics} we present the results of the kinematic analysis as a function of magnitude for the RC, RGBB and AGBB sample and in section \ref{sec:conclusions} we summarise the main conclusions of this work.

%% file: sections/b_KinematicsVVVpm.tex
\begin{figure*}
	\includegraphics[width=\textwidth]{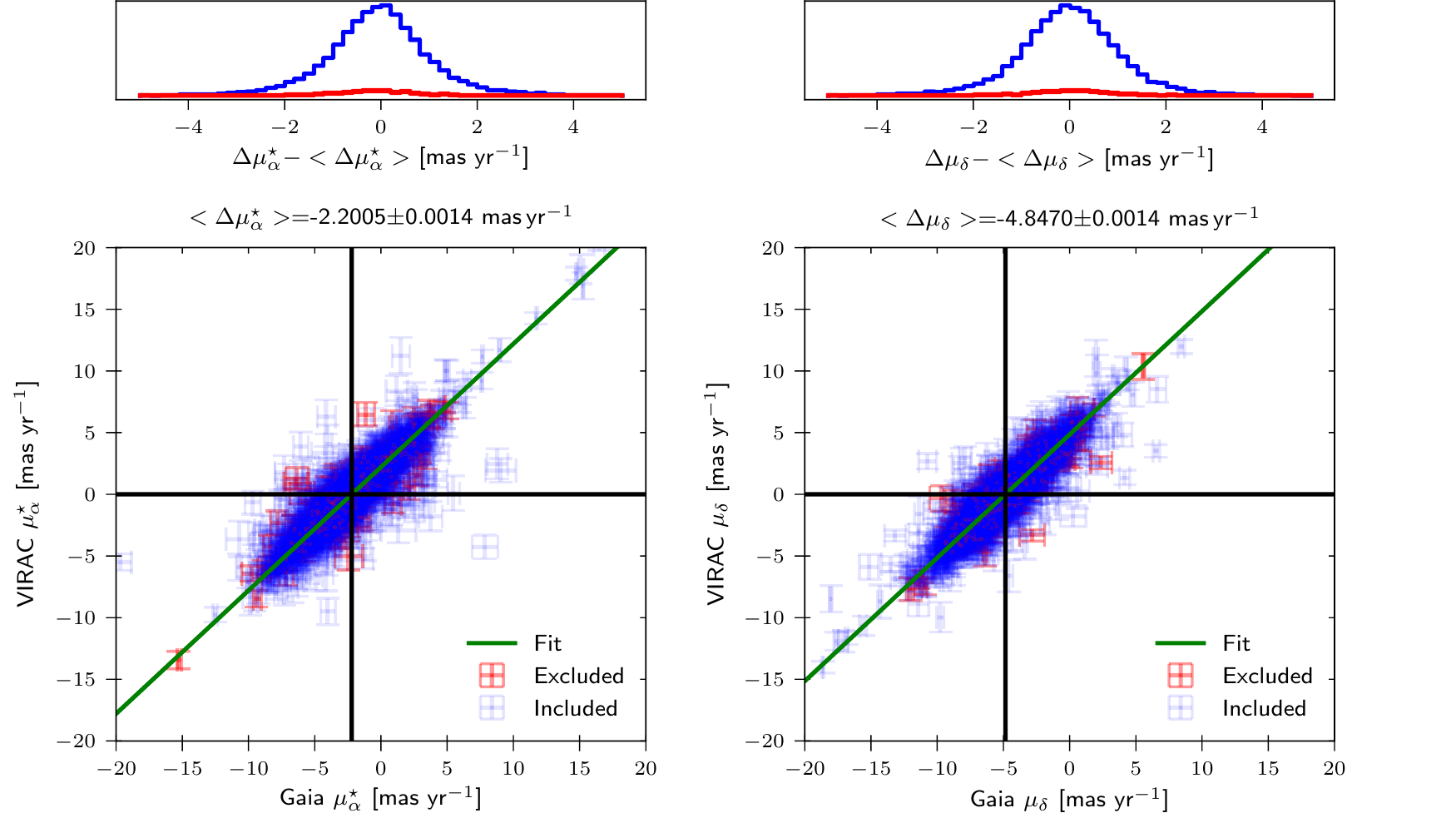}
    \caption{ Tile b278 (1.$^\circ$, -4.2$^\circ$). Comparison between proper motions in RA and DEC measured by \textit{Gaia} and VIRAC. Cross matching performed using a 1." matching radius.
    The bottom row shows the raw proper motion measurements for \textit{Gaia} and VIRAC in RA (left) and DEC (right).
    The blue points are the stars selected for the offset fitting based upon their proper motions errors and other criteria described in the text. 
    The red points were excluded upon application of these criteria.
    There is a linear relationship in both cases, with gradient of 1 by construction, which is shown here as the green line. The black lines show the zero point for the VIRAC proper motions in the \textit{Gaia} reference frame. The mean offset is shown in the plot titles and demonstrates that statistically the mean offset is very well determined due to the large number of stars per tile.
    The top row shows histograms of the deviation from the mean offset of the proper motion difference of individual stars.}
    \label{fig:absolutePM}
\end{figure*}

\begin{figure*}
	\includegraphics[width=\textwidth]{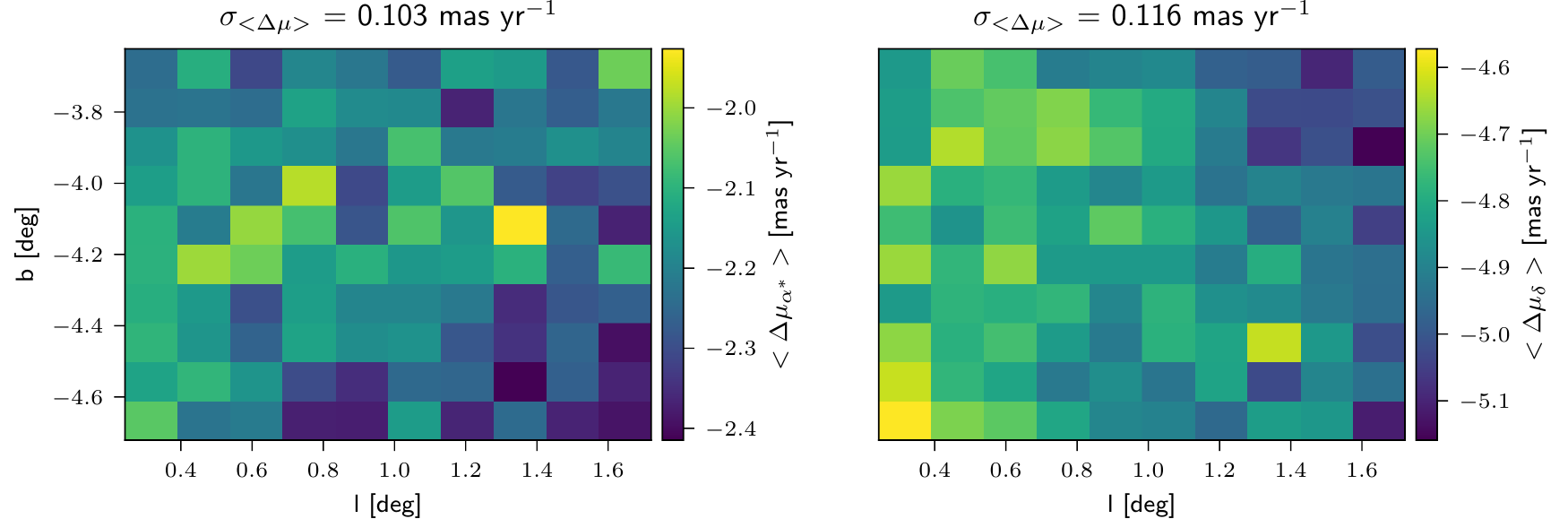}
    \caption{Tile b278 (1.$^\circ$, -4.2$^\circ$). Offsets calculated on a sub tile grid for RA (left) and DEC (right). These maps show there is significant variation of the measured proper motion offset within a tile. The standard deviation of the offsets, see figure titles, is of order 0.1 $\masyr$ and we observe a slight gradient across the map for $<\Delta\mu_\delta>$. These demonstrate that there are systematic effects occurring in the proper motion correction which are likely due to a combination of 
    (i) the known systematics in the \textit{Gaia} proper motion reference frame \citep{lindegren_2018}; and
    (ii) variations in $\mpml$ and $\mpmb$ due to a varying distance distribution of reference sources because of variable extinction.
    }
    \label{fig:absolutePM_offsetTest}
\end{figure*}

\subsection{The VIRAC Proper Motion Catalogue}\label{subsec:virac}

The VISTA Variables in the Via Lactea (VVV) \citep{minniti_2010} survey is a public, ESO, near-InfraRed (IR) survey which scanned the MW bulge, and an adjacent section of the disc at $l<0\dg$. Using the 4m class VISTA telescope for a 5 year period, a typical VVV tile was observed in between 50 to 80 epochs from 2010 to 2015. An extended area of the same region of the galaxy is currently being surveyed as part of the VVVX survey.
The VISTA Infrared Camera (VIRCAM) has a total viewing area of 0.6 deg$^2$ for each pointing with each pointing known as a pawprint. A VVV tile consists of 6 pawprints, three in $l$ times two in $b$, with a total coverage of $\approx$ 1.4 by 1.1$^\circ$, and substantial overlap between the individual pawprints. This overlap ensures that a large number of sources are observed in two or more pawprints.
The bulge region observations are comprised of 196 tiles spanning roughly $-10<l<10^\circ$ and $-10<b<5^\circ$.

The VVV Infrared Astrometric Catalogue (VIRAC) takes advantage of the excellent astrometric capabilities of the VVV survey to present 312,587,642 unique proper motions spread over 560 $\mathrm{deg^2}$ of the MW bulge and southern disc \citepalias{smith_2018}. 
In the astrometric analysis a pawprint set was constructed by cross-matching the telescope pointing coordinates within a 20" matching radius which results in a sequence of images of the same on-sky region at different epochs. Each pawprint set was treated independently to allow precise photometry.
This yielded a total of 2100 pawprint sets from which independent proper motions could be calculated. 
In section 2 of \citetalias{smith_2018} the criteria for rejecting a pawprint are outlined.
Within each pawprint set a pool of reference sources with $\mul$ and $\mub$ not significantly deviant from the local $\mpml$ and $\mpmb$ are extracted in an iterative process. All proper motions within a pawprint set are calculated \textit{relative} to this pool but, because absolute $\mpml$ and $\mpmb$ are unknown at this stage, there is an unknown drift in $l$ and $b$ for each pawprint which we measure in section \ref{subsec:2absPM} using \textit{Gaia} data.
The difference in drift velocity of the reference sources between pawprint sets, within a VVV tile, is smaller than the measurement error on the proper motion measurements from a single pawprint set.
A VVV tile can therefore be considered to be in a single consistent reference frame with a constant offset from the absolute reference frame. 
To calculate final proper motions for stars observed in multiple pawprints \citetalias{smith_2018} use inverse variance weighting of the individual pawprint measurements. 
Also provided is a reliability flag to allow selection of the most reliable proper motion measurements. 
The approach and criteria to determine this flag is presented in section 4.2 of \citetalias{smith_2018}.
In this paper we only use the stars where the reliability flag is equal to one denoting that the proper motion are the most trustworthy.

In this work we adopt the VVV tiling structure for the spatial binning. For integrated on-sky maps we split each tile into quarters for greater spatial resolution. However when considering the kinematics as a function of magnitude we use the full tile to maintain good statistics in each magnitude interval.
For the majority of tiles in the VIRAC catalogue there is photometry in $K_s$, H and J bands. The exceptions are fields b274 and b280 for which VIRAC has no H band data and b212 and b388 for which VIRAC has no J band data.
These data were not present in VVV DR4 when the photometry was added to VIRAC.
We make use of an example tile in figures illustrating the analysis approach. The tile is b278 which is centred at approximately $l$=1.0$\dg$, $b$=-4.2$\dg$.

\subsection{Correction to absolute Proper motions with Gaia}\label{subsec:2absPM}

The VIRAC catalogue presents the proper motions in right ascension (RA), $\mathrm{\mu_{\alpha^*}}$, and declination (DEC), $\mathrm{\mu_{\delta}}$, relative to the mean proper motions in a VVV tile. 
To obtain the absolute proper motions each VVV tile is cross matched with the $Gaia$ DR2 catalogue to make use of its exquisite absolute reference frame \citep{lindegren_2018}. 
Only matches within 1.0 arcsec are considered.

Figure \ref{fig:absolutePM} shows the proper motions as measured by $Gaia$ plotted against the proper motions as measured by VIRAC for VVV tile b278. 
The left panel shows the comparison for RA and the right panel shows the comparison for DEC.
Stars are selected for use in the fitting based upon a series of quality cuts:
\begin{inparaenum}
\item The uncertainty in proper motion measurement is less than 1.5 $\masyr$ for both \textit{Gaia} and VIRAC.
\item The star has an extincted magnitude in the range $10<K_s<15$ mag.
\item The star is classed as reliable according to the VIRAC flag.
\item The cross match angular distance between VIRAC and \textit{Gaia} is less than 0.25".
\end{inparaenum}
These criteria result in a sample of stars for which the mean G band magnitude is $\approx16.5$ with a dispersion of $\approx1.0$ magnitudes. 
By construction a linear relationship, with gradient equal to one, is fit to the distribution. This fits well given that we expect there should be a single offset between \textit{Gaia} and VVV proper motions for each pawprint set.
The offset between the zero point for VIRAC and \textit{Gaia} is caused by the drift motion of the pool of reference stars used for each pawprint set. The measured offsets and uncertainties for the example tile are quoted in figure \ref{fig:absolutePM}.
The consistency checks performed by \citetalias{smith_2018} showed that measurements between different pawprint sets are consistent at the tile scale. A single offset per tile is therefore used to correct from relative proper motions to the absolute frame.

To check this assumption further we computed the offsets on a sub tile scale for tile b278, see figure \ref{fig:absolutePM_offsetTest}. We use a ten by ten sub-grid and determine $\sigma_{\Delta\mu_{\alpha}}$=0.10 $\masyr$ and $\sigma_{\Delta\mu_\delta}$=0.12 $\masyr$. These values show that the uncertainty in the fitted offset is larger than the formal statistical uncertainty derived on the offsets by about two orders of magnitude. We also see indications of a gradient across the tile for the DEC offsets.
These are likely a combination of two effects.
There are known systematics in the \textit{Gaia} proper motion reference frame \citep{lindegren_2018}, an example of which was observed in the LMC \citep{helmi_2018}.
Additionally there are possible variations in $\mpml$ and $\mpmb$ on this scale due to variation in the average distance of the reference sources, causing a variation in the measured mean proper motions, caused by variable extinction.

\subsection{Extracting Red Giants}\label{subsec:getRGonly}

\begin{figure}
	\includegraphics[width=\columnwidth]{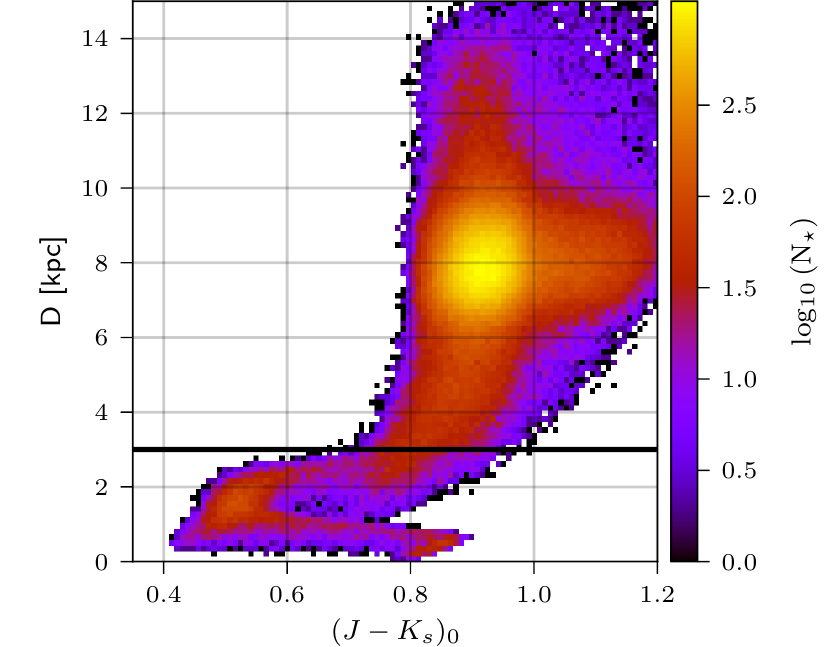}
    \caption{Tile b278 (1.$^\circ$, -4.2$^\circ$). Colour-distance distribution for a single line of sight, and in the magnitude range $11.0<K_{s0}<14.4$ mag, made using the $galaxia$ model. We see a clear MS and then a RG branch with a strong density peak at the galactic centre, much of which is due to RC stars at this distance.
    The RG stars are clearly separated spatially from the MS stars that can only be observed when at distances D $\lesssim$ 3 kpc (horizontal black line). We remove the FG MS stars as they will have disc kinematics and we wish to study the kinematic structure of the bulge-bar.}
    \label{fig:galaxiaDistanceMap}
\end{figure}

The stellar population observed by the VVV survey can be split into two broad categories; the foreground (FG) disk stars and the bulge stars. 
Figure \ref{fig:galaxiaDistanceMap} shows the colour-distance distribution of a stellar population model made using $galaxia$ \citep{sharma_2011}. 
The model was observed in a region comparable to the example tile and only stars with $K_{s0}<14.4$ mag are used.
The FG disk stars are defined to be those that reside between the bulge and the Sun, at distances D $\lesssim$ 4 kpc. 
Considering the magnitude range $11.5<K_{s0}<14.4$ mag we work in, the stars observed at D $\lesssim4$ kpc will be mostly main sequence (MS) stars. 
The bulge stars residing at distances D $>$ 4 kpc are expected to be predominantly RG stars.
Figure \ref{fig:galaxiaDistanceMap} is analogous to a colour-absolute magnitude diagram and shows the two stellar types are separated spatially along the line of sight with only a relatively small number of sub-giant (SG) stars bridging the gap. 

\begin{figure}
	\includegraphics[width=\columnwidth]{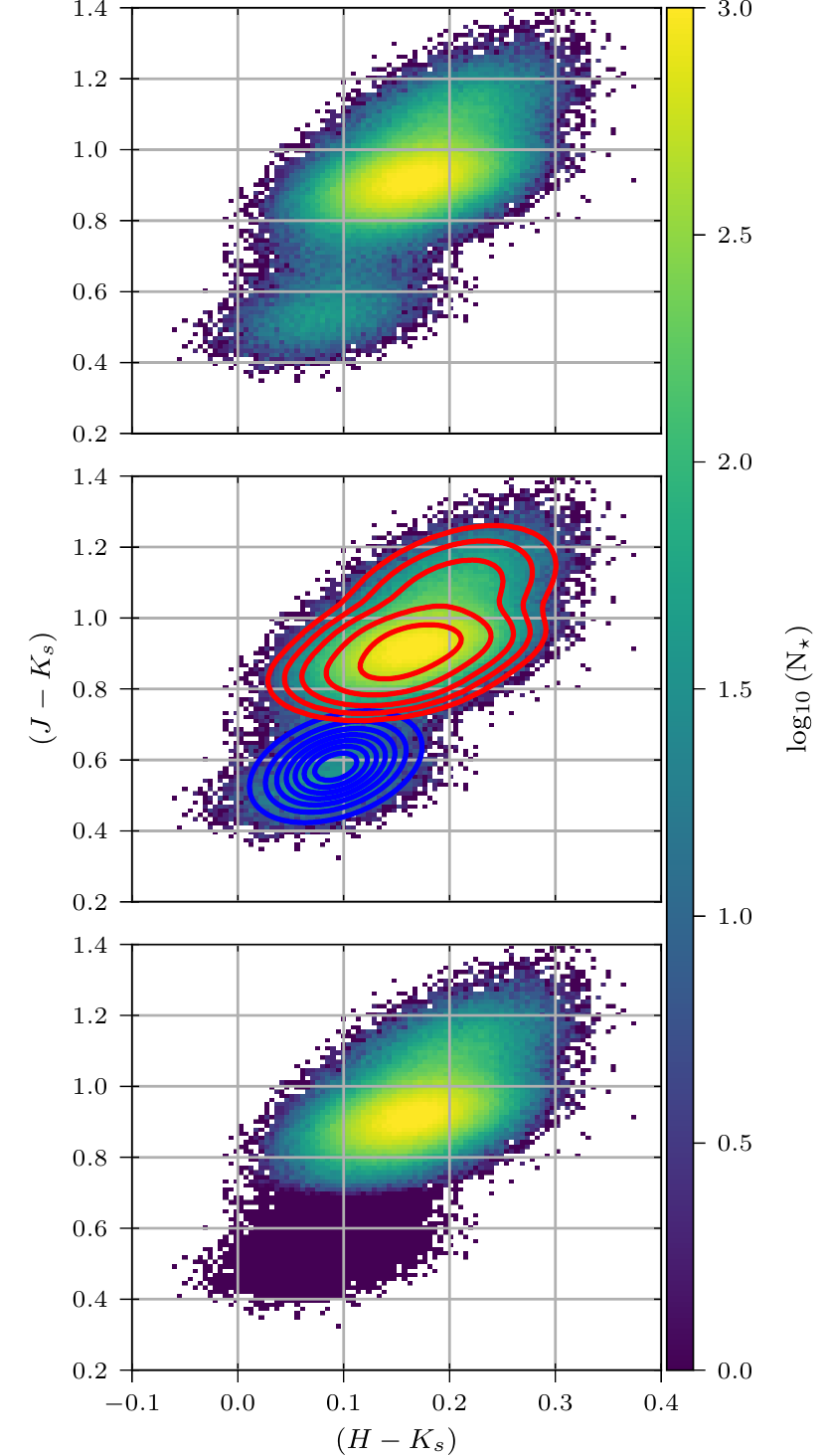}
    \caption{Tile b278 (1.$^\circ$, -4.2$^\circ$). Illustration of the colour selection procedure for the \textit{galaxia} synthetic stellar population. The top panel shows the reddened colour-colour log density diagram for the example tile. The middle panel shows the gaussian mixtures that have been fitted to this distribution. The blue contours highlight the foreground population and the red contours show the RGs.
    The bottom panel shows the RGB population following the subtraction of the FG component.}
    \label{fig:colorcolor_galaxia}
\end{figure}

To study the kinematics of the bulge we remove the FG stars to prevent them contaminating the kinematics of the bulge stars. 
Considering the colour-colour distribution of stars, $(J-K_s)$ vs $(H-K_s)$, we expect the bluer FG to separate from the redder RG stars, see figure \ref{fig:galaxiaDistanceMap}. 
We use the colour-colour distribution as the stars' colours are unaffected by distance. 
A stellar population that is well spread in distance will still have a compact colour-colour distribution if the effects of extinction and measurement uncertainties are not too large.
The top panel of figure \ref{fig:colorcolor_galaxia} shows the colour-colour distribution for the $galaxia$ model observed in the example tile. 
There are two distinct features in this diagram. 
The most apparent feature is the redder (upper right) density peak that corresponds to stars on the RGB.
The second feature is a weaker, bluer density peak (lower left) which corresponds to the MS stars. 
These two features overlap due to the presence of sub-giants which bridge the separation in colour-colour space. 
In tiles where there is more extinction the RGB component is shifted to even redder colours. 
The MS stars, which are closer, are not obscured by the extinction to the same extent and are not shifted as much as the RG stars.
This increases the distinction between the two components and so we separate based upon colour before correcting for extinction.

\begin{figure}
	\includegraphics[width=\columnwidth]{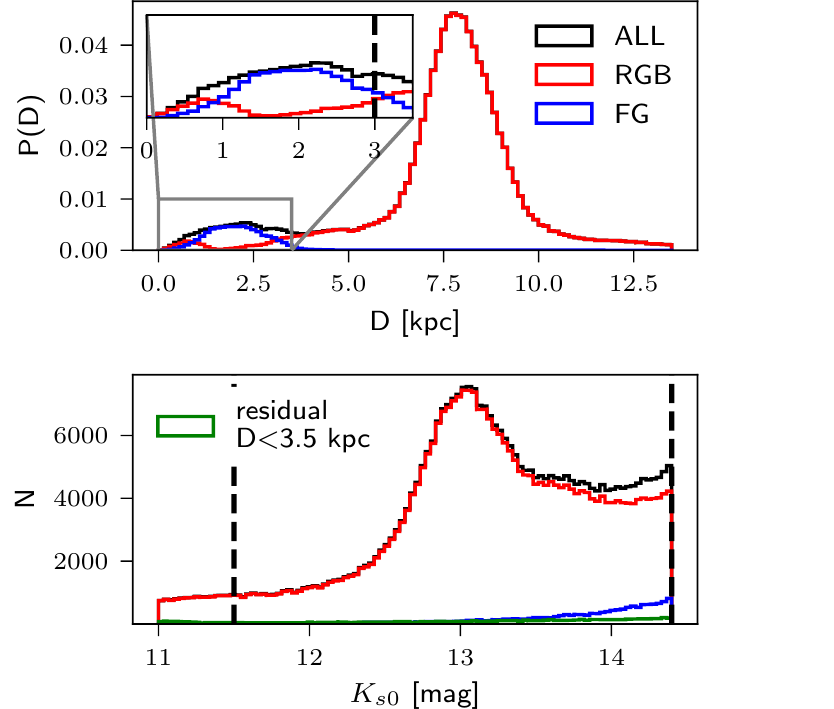}
    \caption{Tile b278 (1.$^\circ$, -4.2$^\circ$). 
    Top panel: Distance distribution of the galaxia synthetic stellar population. The whole distribution is outlined in black and the sample has been divided according to the result of the GMM fitting for the foreground. The stars called RGB are shown in red and the FG component in blue. 
    We zoom in on the 0. $<$ D/kpc $<$ 3.5 region of the plot to provide greater clarity.
    Bottom panel: The same decomposition now mapped into magnitudes. In addition we show the contribution of the stars classed as RGB by the GMM that are at distances D $<$ 3.5 kpc as the green histogram. These stars contribute $\sim0.6\%$ of the total RGB population.
    This shows that the GMM modelling is successful in identifying most of the MS foreground stars with only a slight residual contamination.   }
    \label{fig:distance_distribution_galaxia}
\end{figure}

\begin{figure}
	\includegraphics[width=\columnwidth]{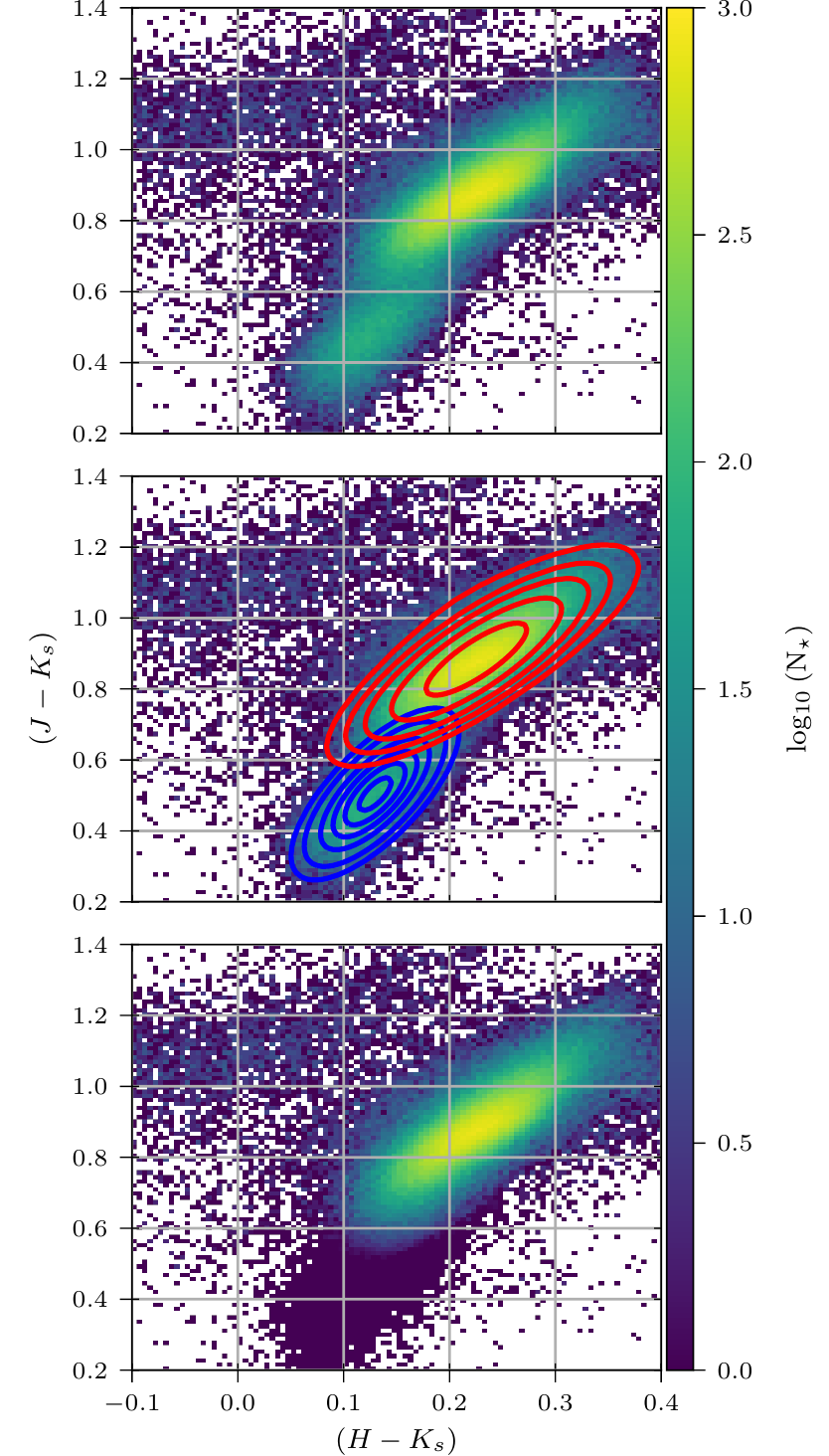}
    \caption{Tile b278 (1.$^\circ$, -4.2$^\circ$). Plots illustrating the separation of FG stars from the RG stars for the VVV example tile using a GMM technique. Top: Colour-colour histogram for the example tile. There are two populations, FG and RGB stars, that overlap slightly in this space but are clearly individually distinct density peaks. Middle: GMM contours showing the fit to the colour-colour distribution. The fit has correctly identified the two populations and allows a probability of the star belonging to either population to be assigned. Bottom: Histogram of the same data where each particle is now weighted by probability of being a RG. 
    The FG component has been successfully removed. There is a smooth transition in the overlap region between FG and RGB with no sharp cutoffs in the number counts of stars. This is expected from a realistic stellar population and cannot be achieved with a simple colour cut.
    }
    \label{fig:colorcolor}
\end{figure}

We use gaussian mixture modelling (GMM) to fit a multi component 2D gaussian mixture (GM) to the colour-colour distribution. 
Fitting was performed with $scikit-learn$ \citep{scikit-learn}. 
The fit is improved by using only stars with an extinction corrected magnitude $K_{s0}<14.4$ mag, see section \ref{subsec:extinction} for details of the extinction correction.
At fainter magnitudes the FG and RGB sequences merge together and it becomes increasingly difficult for the GMM to accurately distinguish the two components.

We use different numbers of gaussians depending on the latitude, and the fits have been visually checked to ensure that they have converged correctly.
Identifying the FG component and the RG component, we weight each star by its probability of being a RG star. 
The weighting is calculated as follows,
\begin{equation}\label{eqn:rcWeighting}
    w_{\mathrm{RG}} = \frac{ P\left( \mathrm{RG} \right) }{P\left( \mathrm{RG} \right) + P\left( \mathrm{FG} \right)},
\end{equation}
where P(RG) and P(FG) are the probability of a star's colours given the RGB and FG gaussian mixtures respectively, and $\mathrm{w_{RG}}$ can take values in the range 0 to 1. 
For the few stars that do not have a measured J band magnitude we assign a weighting equal to one. 
These stars are mostly highly reddened, causing their J band magnitude to not be measured and are therefore likely to be bona fide bulge stars.
To test the procedure outlined above it was applied to the $galaxia$ model. The model has had extinction applied and the magnitudes are randomly convolved with typical observational uncertainties to mimic the VVV survey. When selecting only the bright stars to apply the modelling we correct the mock extincted magnitudes using the same method as is used on the data to make the test as consistent as possible. 
The progression is shown in figure \ref{fig:colorcolor_galaxia} with the top panel outlining the double peaked nature of the colour-colour diagram. The middle panel shows the fitted gaussians, FG in blue and RGB in red, and the bottom panel showing the original histogram now weighted according to equation \ref{eqn:rcWeighting}. The GMM has identified the density peaks correctly and removed the stars in the FG part of the diagram.
Figure \ref{fig:distance_distribution_galaxia} shows the results of the GMM procedure on the $galaxia$ population's distance (top) and luminosity function (bottom). The GMM successfully removes the majority of stars at distances $D<3$ kpc. The contamination fraction in the RGB population by stars at $D<5$ kpc distance is then only $\approx 1\%$. 
Figure \ref{fig:distance_distribution_galaxia} also shows the presence of a FG population that corresponds to the blue MS population shown in figure \ref{fig:galaxiaDistanceMap} at colours $(J-K_s)_0\lesssim0.7$. At D$\lesssim$1.2 kpc a small number of stars are included in the RGB population which plausibly correspond to the redder faint MS population seen in figure \ref{fig:galaxiaDistanceMap}. This population accounts for $\sim0.6\%$ of the overall RGB population. The RGB population tail at D $\lesssim$ 3 kpc is composed of SG stars. The GMM is clearly extremely successful at removing the MS stars and leaving a clean sample of RGB with a tail of SG stars.

Having demonstrated that the GMM colour selection process works we apply it to each tile. Figure \ref{fig:colorcolor} shows the progression for tile b278. This plot is very similar to figure \ref{fig:colorcolor_galaxia} and gives us confidence that the GMM procedure is a valid method to select the RGB bulge stars. 
The sources at low $(H-K_s)$ and high $(J-K_s)$ present in the data but not the model are low in number count and do not comprise a significant population.

\begin{figure}
	\includegraphics[width=\columnwidth]{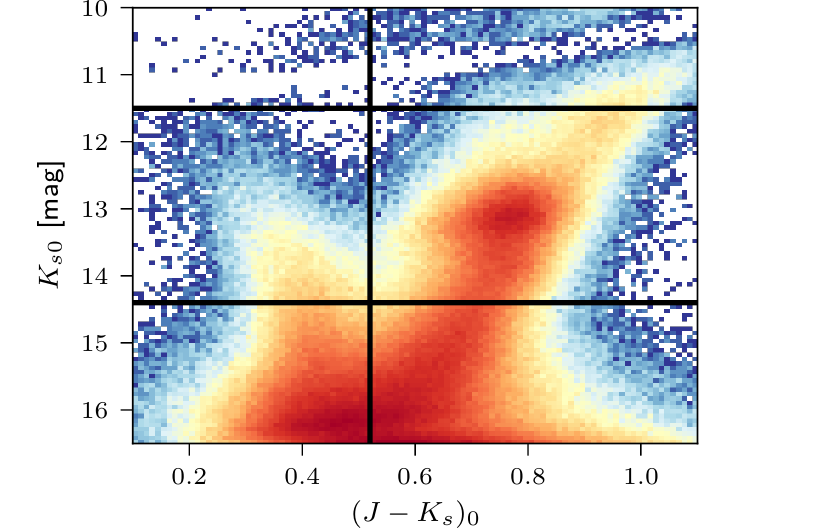}
    \caption{ Tile b274 (-4.8$^\circ$, -4.2$^\circ$). Colour magnitude diagram for one of two tiles with no H band observations and requiring a colour cut at $(J-K_s)_0=0.52$ mag (vertical black line) to separate the FG stars.
    The two horizontal lines mark the boundary of our magnitude range of interest at $11.5<K_{s0}<14.4$ mag. The fainter boundary is selected to be brighter than where the FG and RGB populations merge in this diagram which aids in the application of the colour-colour selection in tiles with full colour information.}
    \label{fig:b274_color_mag}
\end{figure}

As mentioned in section \ref{subsec:virac} there are 4 tiles with incomplete observations in either H or J bands. Tiles b274 and b280 have no H band measurements in VIRAC and the colour-colour approach cannot be applied. For these tiles we apply a standard colour cut at $(J-K_s)_0<0.52$ to remove the FG stars. Figure \ref{fig:b274_color_mag} illustrates this cut and also includes lines highlighting the magnitude range we work in, $11.5< K_{s0}< 14.4$ mag. The fainter limit is at the boundary where the FG and RGB sequences are beginning to merge together and the brighter limit is fainter than the clear artefact which is likely due to the VVV saturation limit.

We exclude the two tiles with no J band observations from the analysis as we do not wish to include the extra contamination due to the foreground in these two tiles. These tiles are plotted in grey throughout the rest of the paper.

\subsection{Extinction Correction}\label{subsec:extinction}

\begin{figure}
	\includegraphics[width=\columnwidth]{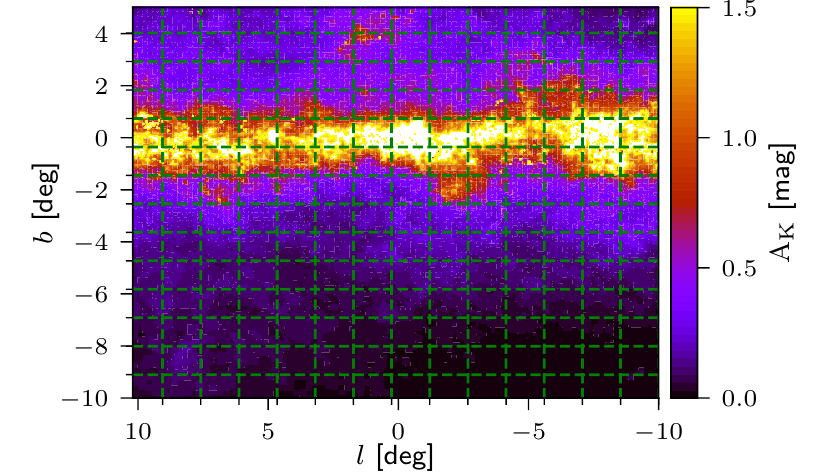}
    \caption{Extinction data from \citet{gonzalez_2012}. Map showing the $K_s$ band extinction coefficient $A_K$ at a resolution of 2'.  It shows the large extinction in the galactic plane and also in places out to $|b|<2^\circ$. Overplotted on this map are the outlines of the VVV tiling pattern with tile b201 at the bottom right, tile b214 at the bottom left and tile b396 at the top left.}
    \label{fig:extinction}
\end{figure}

By observing in the IR, VVV can observe a lot deeper near the galactic plane where optical instruments like \textit{Gaia} are hindered by the dust extinction. 
However, at latitudes $|b|<2^{\circ}$ the extinction becomes significant even in the IR, with $A_K > 0.5$. 
We use the extinction map derived by \citet{gonzalez_2012}, shown together with the VVV tile boundaries in figure \ref{fig:extinction}, to correct the $K_s$ band magnitudes directly following $K_{s}=K_{s0}+A_K(l,b)$ where $K_{s0}$ is the unextincted magnitude. This map has a resolution of 2'.
We correct H and J bands, where available using the $A_K$ values from the map and the coefficients $A_H/A_K=1.73$ and $A_J/A_K=3.02$ \citep{nishiyama_2009}.
We use the extinction map as opposed to an extinction law because some of the stars do not have the required H or J band magnitudes.

A further issue, caused partially by extinction but also by crowding in the regions of highest stellar density, is the incompleteness of the VVV tiles. 
Our tests have demonstrated that at latitudes $|b|>1.0\dg$ and away from the galactic centre, ($|l|>2.0\dg$,$|b|>2.0\dg$), the completeness is $>80\%$ at $K_{s0}= 14.1$ mag. However inside these regions the completeness is lower, and so we exclude these region from our magnitude dependant analysis.

Our extinction correction assumes that the dust is a foreground screen. Due to the limited scale height of the dust this is a good assumption at high latitude. The assumption becomes progressively worse at lower latitudes and the distribution of actual extinctions increasingly spreads around the map value due to the distance distribution along the line of sight.
Due to incompleteness we exclude the galactic plane, which is also where the 2D dust assumption is worst, from our magnitude dependent analysis.
We further apply a mask at $A_K=1.0$ mag when considering integrated on-sky maps.

%% file: sections/c_KinematicsNmagic.tex
We compare the VIRAC proper motions to the MW bar models of \citetalias{portail_2017}. They used the made-to-measure (M2M) method to adapt dynamical models to fit the following constraints:
\begin{inparaenum}
    \item The RC density computed by \citetalias{wegg_2013} by inverting VVV star count data.
    \item The magnitude distributions in the long bar from UKIDSS and 2MASS surveys \citepalias{wegg_2015}.
    \item The stellar kinematics of the BRAVA  \citep{howard_2008,kunder_2012} and ARGOS \citep{freeman_2013,ness_2013} surveys.
\end{inparaenum}
The models very successfully reproduce the observed star counts and kinematics for pattern speeds in the range 35.0$<\Omega<$42.5 $\kmskpc$. \citetalias{portail_2017} found a best fitting bar pattern speed of 39.0$\pm$3.5$\kms$ however in this work we use the model with $\Omega=37.5$ $\kmskpc$ together with a slightly reduced total solar tangential velocity $V_{\phi,\odot}=245$ $\kms$ as we see an improved match between the $\mpml$ maps. 
In the integrated maps, see section \ref{sec:rg_kinematics}, the shape of the $\mpml$ isocontours is improved. In the magnitude sliced maps, see section \ref{sec:rc_kinematics}, the gradient between bright and faint magnitude is better reproduced by this model. 
In future work we shall explore quantitatively the constraints on the pattern speed, solar velocity and mass distribution that can be obtained from VIRAC.
The other solar velocities remain unchanged from \citetalias{portail_2017}; we use a radial solar velocity $V_{r,\odot}=-11.1$ $\kms$ (i.e. moving towards the GC), and a vertical solar velocity of $V_{z,\odot}=7.25$ $\kms$ \citep{schoenrich_2010}. Our chosen fiducial barred model has a mass-to-clump ratio (the total mass of the stellar population, in $M_\odot$, that can be inferred from the presence of one RC star) of 1000, and a nuclear stellar disc mass of 2.0$\times$10$^9$ $M_\odot$, see \citetalias{portail_2017}.

The aim of this section is to construct a model stellar distribution with magnitude and velocity distributions that can be directly compared to VIRAC. The \citetalias{portail_2017} model provides the kinematics and the distance moduli of the particles. The distance moduli are calculated assuming $R_o=8.2$ kpc \citep{bland_hawthorn_2016} which is very similar to the recent GRAVITY results \citep{gravity_2019}. To construct the magnitude distribution we further require an absolute luminosity function (LF) representing the bulge stellar population and we use the distance moduli to shift this LF to apparent magnitudes.
Each particle in the model can be thought of as representing a stellar population with identical kinematics.

\subsection{Synthetic Luminosity Function}\label{subsec:particle2stellarDist}

To construct an absolute LF representing the bulge stellar population we used:
\begin{inparaenum}
\item The Kroupa initial mass function \citep{kroupa_2001} as measured in the bulge \citep{wegg_2017};
\item a kernel-smoothed metallicity distribution in Baade's window from \citet{zoccali_2008} where we use the metallicity measurement uncertainty to define each kernel;
\item isochrones describing the stellar evolution for stars of different masses and metallicities.
\end{inparaenum}
The PARSEC + COLIBRI isochrones \citep{bressan_2012,marigo_2017} were used with the assumption that the entire bulge population has an age of 10 Gyr \citep{clarkson_2008,surot_2019}. 
These three ingredients were combined in a Monte Carlo simulation where an initial mass and metallicity are randomly drawn and then used to locate the 4 nearest points on the isochrones. 
Interpolating between these points allows the [$M_K$,$M_H$,$M_J$] magnitudes of the simulated star to be extracted. 
The simulation was run until $\mathrm{10^6}$ synthetic stars had been produced.

\begin{figure}
	\includegraphics[width=\columnwidth]{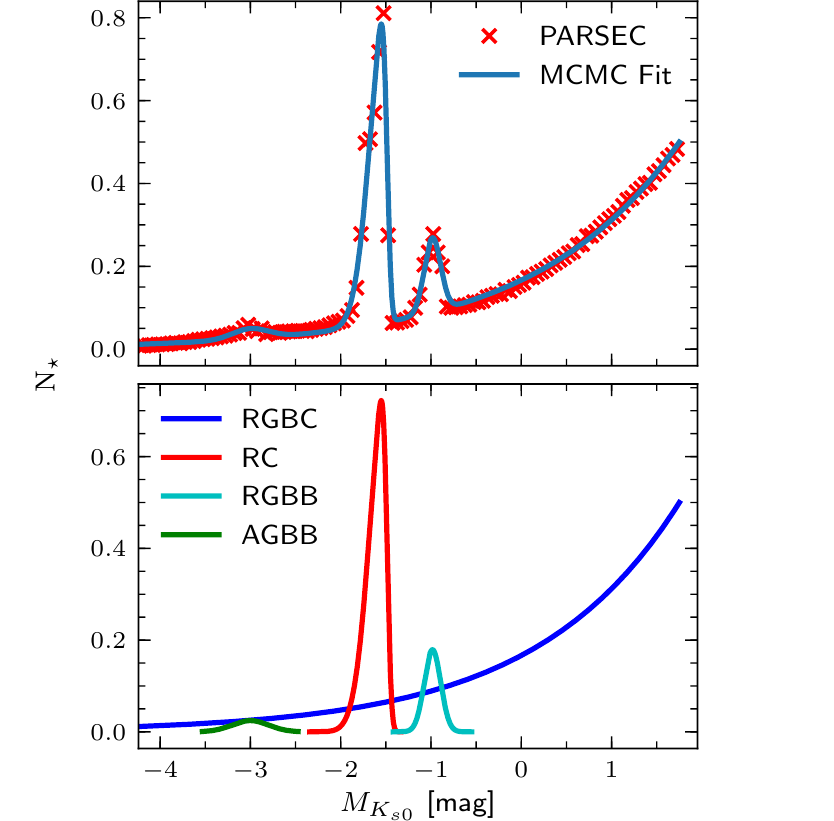}
    \caption{
    Theoretical luminosity function used as inputs to the modelling to facilitate the observation of the particle model consistently with the VVV survey. 
    Top: 
    The initial LF is shown in red crosses. This is produced from the Monte Carlo sampling and the colour-colour selection procedure has been applied in a manner consistent with the VIRAC data.
    The Markov Chain Monte Carlo fit using four components, an exponential background, a gaussian each for the AGBB and RGBB, and a skewed gaussian for the RC, is overplotted as the blue line.
    Bottom: LF now split into the components that will be used in this paper; the RC (red), RGBB (cyan), AGBB (green), that are combined to produce the RC\&B, and the RGBC (blue). }
    \label{fig:lf}
\end{figure}

To observe the model as if it were the VIRAC survey it is necessary to implement all the associated selection effects. 
In section \ref{subsec:getRGonly} a colour based selection was used to weight stars based on their probability of belonging to the RGB. 
The same colour based procedure was applied to the synthetic stars' colour-colour diagram and the corresponding weighting factors were calculated. 
The results of the simulation, with the colour weightings applied, are shown in the upper panel of figure \ref{fig:lf}. 
As expected, the RC LF is very narrow facilitating their use as standard candles in studies of the MW (eg. \citealt{stanek_1994,bovy_2014,wegg_2015}).

\begin{table}
	\centering
	\caption{
	Reference table of the most commonly used acronyms.
	}
	\label{tab:acronyms}
	\begin{tabular}{cl} 
		\hline
		Acronym & Definition\\
		\hline
		LF & Luminosity Function\\
		FG & Foreground\\
		SG & Sub-Giant\\
		RGB & Red Giant Branch\\
		RC & Red Clump\\
		RGBB & Red Giant Branch Bump\\
		AGBB & Asymptotic Giant Branch Bump\\
		RGBC & Red Giant Branch Continuum\\
		RC\&B & Red Clump and Bumps\\
		\hline
	\end{tabular}
\end{table}


We define the exponential continuum of RGB stars, not including the over densities at the RC, RGBB and AGBB, to be a distinct stellar population, henceforth referred to as the red giant branch continuum (RGBC). 
We refer to the combined distribution of the RC, RGBB and AGBB stars as the RC\&B.
A list of stellar type acronyms used in this paper is given in table \ref{tab:acronyms}.

We fit the simulated LF with a four component model that we then combine to construct the RGBC and RC\&B. We use an exponential for the RGBC,
\begin{equation}\label{eqn:rgbc_lf}
    \Lagr_{\mathrm{RGBC}}\left(\mk\right) = \alpha  \exp{\left( \beta  M_{K_{s0}} \right)}.
\end{equation}
We fit separate gaussians for the RGBB and AGBB,
\begin{equation}
    \Lagr_{\mathrm{RGBB/AGBB}}\left(\mk\right) = \frac{C_i}{\sqrt{2\pi\sigma_i^2}} \exp{\left(-\frac{1}{2} \zeta_i^2 \right) },
\end{equation}
where,
\begin{equation}
    \zeta_i=\frac{ M_{K_{s0}} - \mu_i }{\sigma_i},
\end{equation}
and $\mu_i$, $\sigma_i$, and $C_i$ denote the mean, dispersion, and amplitude of the respective gaussians. 
We use a skewed gaussian for the RC distribution,
\begin{equation}
    \Lagr_{\mathrm{RC}}\left(\mk\right) = \frac{C_{RC}}{\sqrt{2\pi\sigma_{RC}^2}}
    \exp{\left(-\frac{1}{2} \zeta_{RC}^2 \right) }
    \left[ 
    1 + \mathrm{erf}\left( \frac{\gamma}{\sqrt{2}}\zeta_{RC} \right)
    \right],
\end{equation}
where $\mathrm{erf}\left(\right)$ is the standard definition of the error function and $\gamma$ is the skewness parameter.
Fitting was performed using a Markov Chain Monte Carlo procedure; the results are shown in the lower panel of figure \ref{fig:lf} and the fitted parameters are presented in table \ref{tab:fitparams}.
These four LFs are used as individual inputs to the modelling code and allow each particle to be observed as any required combination of the defined stellar evolutionary stages.
These choices are well motivated as \citet{nataf_2010} and \citetalias{wegg_2013} showed that the RGBC is well described by an exponential function and the RC LF is known to be skewed \citep{girardi_2016}.


Ideally we would use only the RC stars from VIRAC when constructing magnitude resolved maps as they have a narrow range of absolute magnitudes and so can be used as a standard candle. 
We statistically subtract, when necessary, the RGBC through fitting an exponential.
As shown in figure \ref{fig:lf} the RC and RGBB are separated by only $\approx0.7$ mag. When convolved with the LOS density distribution these peaks overlap.
Because it is difficult to distinguish the RGBB from the RC observationally we accept these stars as contamination. 
It is also important to include the AGBB \citep{gallart_1998}; stars of this stellar type residing in the high density bulge region can make a significant kinematic contribution at bright magnitudes, $K_{s0}<12.5$ mag, where the local stellar density is relatively smaller.

\begin{table}
	\centering
	\caption{
	Parameters for the LF shown in figure \ref{fig:lf}.
	}
	\label{tab:fitparams}
	\begin{tabular}{lr} 
		\hline
		Parameter & Value \\
		\hline
		$\alpha$ & 0.1664 \\
		$\beta$  & 0.6284 \\
		
		$\mu_{\mathrm{RGBB}}$ & -0.9834 \\
		$\sigma_{\mathrm{RGBB}}$ & 0.0908 \\
		$C_{\mathrm{RGBB}}$ & 0.0408 \\
		
		$\mu_{\mathrm{AGBB}}$ & -3.0020 \\
		$\sigma_{\mathrm{AGBB}}$ & 0.2003 \\
		$C_{\mathrm{AGBB}}$ & 0.0124 \\
		
		$\mu_{\mathrm{RC}}$  & -1.4850 \\
		$\sigma_{\mathrm{RC}}$ & 0.1781 \\
		$C_{\mathrm{RC}}$ & 0.1785 \\
		$\gamma$ & -4.9766 \\
		\hline
	\end{tabular}
\end{table}

\subsection{VIRAC Observables}\label{subsec:nmagicObservables}

The kinematic moments we consider are the mean proper motions, the corresponding dispersions and the correlation between the proper motions.

We here define dispersion,
\begin{equation}
    \sigma_{\mu_i} = \sqrt{ <\mu_i^2> - <\mu_i>^2 },
\end{equation}
with $i \in \left( l,b \right)$ and the correlation,
\begin{align}\label{eqn:correlation}
    \mathrm{corr} \left( \mu_l , \mu_b \right) & = 
    \frac{<\mu_l \mu_b > - <\mu_l><\mu_b>}{ \sqrt{ \left( <\mu_l^2> - <\mu_l>^2 \right) \left( <\mu_b^2> - <\mu_b>^2 \right) } }\\
    & = \frac{\sigma_{lb}^2}{\sigma_l\sigma_b}.
\end{align}

In the previous section we described the method to construct synthetic absolute LFs for the RGBC and the RC\&B stars, see figure \ref{fig:lf}. We now combine this with the dynamical model of \citetalias{portail_2017} to observe the model through the selection function of the VIRAC survey. For a more detailed description of the process used to reconstruct surveys see \citetalias{portail_2017}.

Each particle in the model has a weight corresponding to its contribution to the overall mass distribution.
When constructing a measurable quantity, or "observable", all particles that instantaneously satisfy the observable's spatial criteria, i.e. being in the correct region in terms of $l$ and $b$, are considered and the particle's weight is used to determine its contribution to the observable. 
In addition to the particle weight there is a second weighting factor, or "kernel", that describes the selection effects of the survey.
The simplest example of an observable is a density measurement for which,
\begin{equation}
    \rho = \Sigma_{i=0}^n w_i K(z_i), 
\end{equation}
where the sum is over all particles, $w_i$ is the weight of the i$^{th}$ particle, $z_i$ is the particle's phase space coordinates and the kernel $K$ determines to what extent the particle contributes to the observable.
To reproduce VIRAC we integrate the apparent LF of the particle within the relevant magnitude interval to determine to what extent a stellar distribution at that distance modulus contributes. 
For the magnitude range $11.8<K_{s0}<13.6$ mag, which we use for constructing integrated kinematic maps, and the stellar population denoted by X, the kernel is given by,
\begin{equation}
    K(z_i) = \delta(z_i) \int_{K_{s0}=11.8}^{K_{s0}=13.6} \Lagr_X(K_{s0}-\mu_i) dK_{s0}
    \label{eqn:kernel}
\end{equation}
where the LF is denoted $\Lagr_X$, the distance modulus of the particle is $\mu_i$, and $\delta(z_i)$ determines whether the star is in a spatially relevant location for the observable.
More complicated observables are measured by combining two or more weighted sums. For example a mean longitudinal proper motion measurement is given by,
\begin{equation}
    \mpml = \frac{\Sigma_{i=0}^n w_i K(z_i) \mu_{l,i}}{\Sigma_{i=0}^n w_i K(z_i)},
    \label{eqn:mean}
\end{equation}
where $\mu_{l,i}$ is the longitudinal proper motion of the i$^{th}$ particle.
This generalises to all further kinematic moments as well. 

To account for the observational errors in the proper motions we input the median proper motion uncertainty measured from the VIRAC data for each tile. We use the median within the integrated magnitude range for the integrated measurements, see section \ref{sec:rg_kinematics}, and the median as a function of magnitude for the magnitude resolved measurements, see section \ref{sec:rc_kinematics}.
Given the true proper motion of a particle in the model we add a random error drawn from a normal distribution centred on zero and with width equal to the median observational error.

Temporal smoothing allows us to reduce the noise in such observables by considering all previous instantaneous measurements weighted exponentially in look-back time \citepalias{portail_2017}.

%% file: sections/d_RGB_kinematics.tex

The methods described in section \ref{sec:vvvpm} were applied to all tiles in VIRAC to extract a sample of stars weighted by their likelihood of belonging to the RGB.
For each quarter tile we implement cuts in proper motion to exclude any high proper motion stars likely to be in the disc and to ensure we only use high quality proper motions: 
We cut all stars with an error in proper motion greater than 2.0 $\masyr$ and apply a sigma clipping algorithm that cuts stars at 3$\sigma$ about the median proper motion. 
There were two stopping criteria; when the change in standard deviation was less than 0.1 $\masyr$ or a maximum of four iterations. These criteria ensure that we only remove the outliers and leave the main distribution unchanged.
These cuts remove $\sim20$\% of the stars in the VIRAC catalogue.
From the resulting sample the on-sky, integrated LOS kinematic moments were calculated, combining the proper motion measurements using inverse variance weighting. 
As discussed in section \ref{subsec:nmagicObservables} we do not remove the additional dispersion caused by measurement uncertainties but instead convolve the model. The typical median error is $\sim1.0$ $\masyr$ which corresponds to dispersion broadening in the range 0.15 to 0.25 $\masyr$.
We note here that there is an uncertainty in the mean proper motion maps of $\sim0.1$ $\masyr$ due to the correction to the absolute reference frame, see section \ref{subsec:2absPM}.
The resulting kinematic maps are compared to the \citetalias{portail_2017} fiducial bar model predictions, as described in section \ref{sec:m2m}.

\begin{figure*}
	\includegraphics[width=\textwidth]{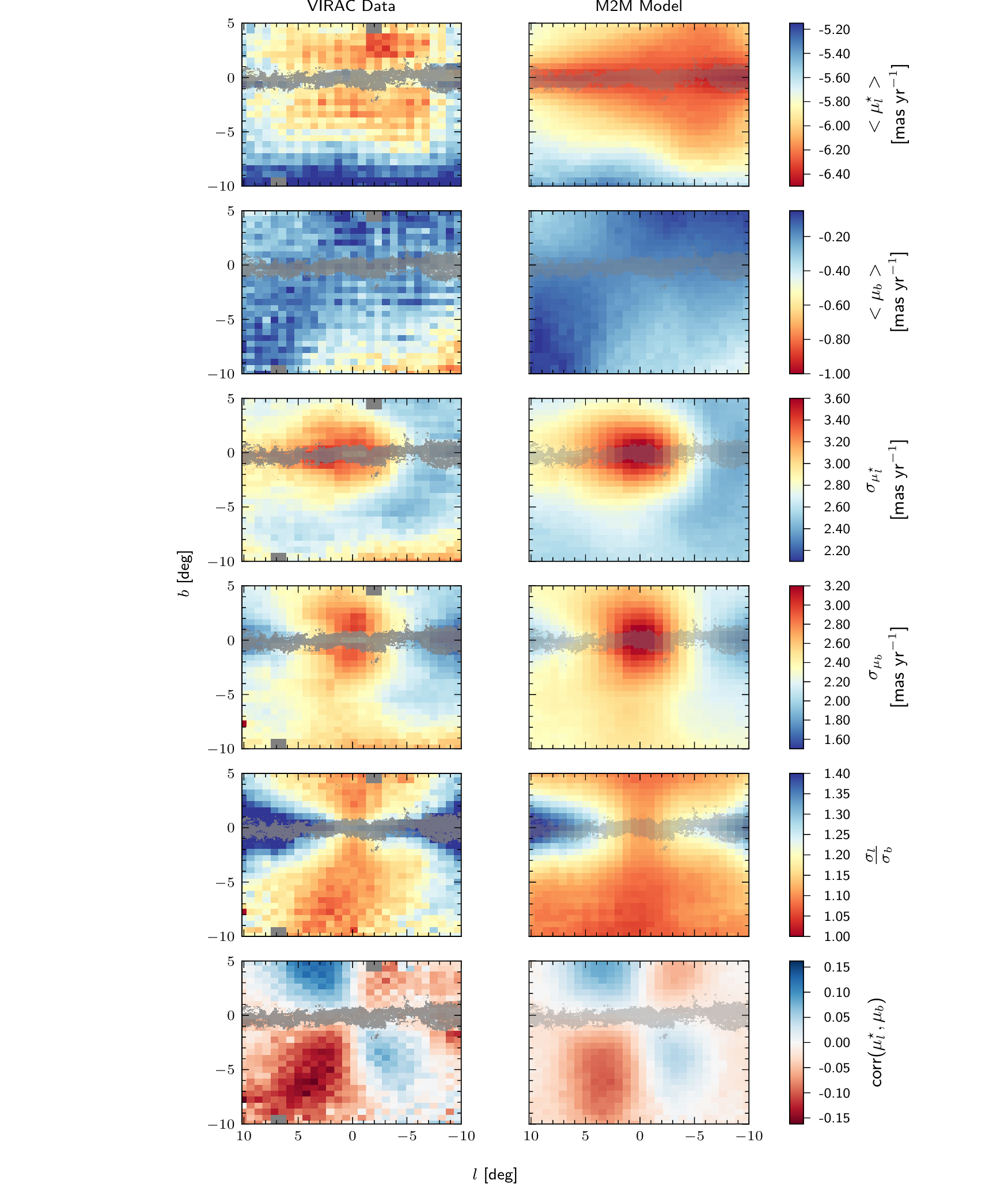}
    \caption{Integrated kinematic maps for the VIRAC data (left column) and the fiducial bar model (right column). The integration magnitude interval is $11.8<K_{s0}<13.6$ mag. The kinematic moments shown are as follows: $\mpml$ and $\mpmb$ (first - second row), $\dpml$, $\dpmb$, dispersion ratio (third - fifth row) and correlation of proper motion vectors (final row).
    The grey mask covers regions for which $A_K>1.0$.
    We see excellent agreement between the model and the data giving us confidence in the barred nature of the bulge.}
    \label{fig:integrated_maps}
\end{figure*}

\begin{figure*}
  \centering
  \includegraphics[width=\textwidth]{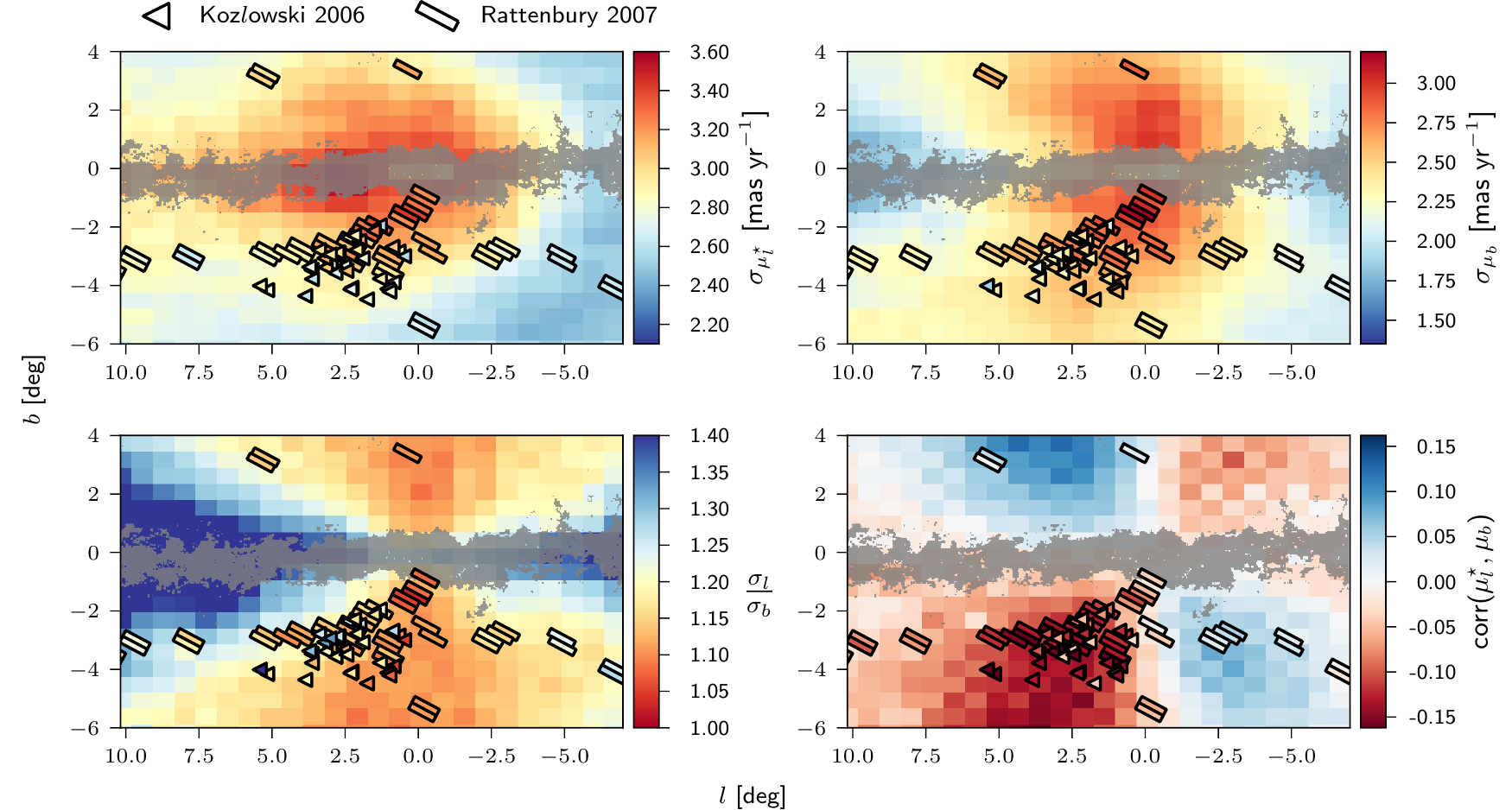}
  \caption{ Comparison between VIRAC proper motions and previous MW bulge proper motion studies (\citetalias{kozlowski_2006} and \citetalias{rattenbury_2007}). The panels show $\dpml$ (top left), $\dpmb$ (top right), $\dpml / \dpmb$ (bottom left) and the correlation (bottom right). In these plots we have zoomed in on the overlap region between the previous datasets and the VIRAC maps.
  The grey mask covers regions for which $A_K>1.0$.}
  \label{fig:compare2otherData}
\end{figure*}

\subsection{Integrated Kinematics For All Giant Stars}
We first present integrated kinematic moments calculated for the magnitude range 11.8 $<K_{s0}<$ 13.6 mag which extends roughly $\pm3$ kpc either side of the galactic centre. Figure \ref{fig:integrated_maps} shows $\mpml$, $\mpmb$, $\dpml$, $\dpmb$, the dispersion ratio, and [$\mul$,$\mub$] correlation components and compares these to equivalent maps for the fiducial model.

The $\mpml$ maps show the projected mean rotation of the bulge stars where the global offset is due to the tangential solar reflex motion measured to be -6.38 $\masyr$ using Sgr A* \citep{reid_2004}.
They contain a clear gradient beyond $|b|>3.\dg$ with the mean becoming more positive at positive $l$ because of the streaming velocity of nearby bar stars, see also section \ref{sec:rc_kinematics}, figure \ref{fig:rcb_meanPM_L_vvv}. 
A similar result was also reported by \citet{qin_2015} from their analysis of an N-body model with an X-shaped bar.
Away from the galactic plane the model reproduces the data well. It successfully reproduces the $\mpml$ isocontours which are angled towards the galactic plane. These isocontours are not a linear function of $l$ and $b$ and have an indent at $l=0\dg$ likely caused by the boxy/peanut shape of the bar.

The $\mpmb$ maps show a shifted quadrupole signature. There are two factors we believe contribute to this effect; the pattern rotation and internal longitudinal streaming motions in the bar. 
The near side of the bar at positive longitude is rotating away from the sun and the far side is rotating towards the sun. The resulting change in on-sky size manifests as $\mub$ proper motions towards the galactic plane at positive longitudes and away from the galactic plane at negative longitudes.
The streaming motion of stars in the bar has a substantial component towards the sun in the near side and away from the sun in the far side which has been seen in RC radial velocities \citep{vasquez_2013}. 
For a constant vertical height above the plane, motion towards the sun will be observed as $+\mub$. 
By removing the effect of the solar motion in the model, and then further removing the pattern rotation, we estimate the relative contribution to $\mpmb$ from the pattern rotation and internal streaming to be 2:1.
The offset of $\approx$ -0.2 $\masyr$ from zero in $\mub$ is due to the solar motion, $V_{z,\odot}$.
The quadrupole signature is also offset from the minor axis due to the geometry at which we view the structure.
It should be noted here that the random noise in the mean proper motion maps is greater than that of the corresponding dispersions. This is a consequence of systematic errors introduced by the \textit{Gaia} reference frame correction \citep{lindegren_2018} to which the mean is more sensitive.

The dispersion maps both show a strong central peak around the galactic centre. This is also seen in the model and is caused by the deep gravitational potential well in the inner bulge. 
In both cases the decline in dispersion away from the plane is more rapid at negative longitude while at positive longitude there are extended arms of high dispersion.  
For both dispersions there is a strip of higher dispersion parallel to the minor axis and offset towards positive longitude; centred at $l\sim1\dg$. This feature is prominent for both data and model for the latitudinal proper motions. For the longitudinal case the model shows this feature more clearly than the data but the feature is less obvious compared to the latitudinal dispersions.
Both maps also show a lobed structure which is also well reproduced by the bar model and is likely a result of the geometry of the bar combined with its superposition with the disc. The model is observed at an angle of 28.0$\dg$ from the bar's major axis \citepalias{portail_2017} and so at negative longitudes the bar is further away and therefore the proper motion dispersions are smaller.
On the other side, for sub-tiles at $l>7.0\dg$ the dispersions are larger and both dispersions decline more slowly moving away from $b=0\dg$, as in this region the nearby side of the bar is prominent. 

The dispersion ratio $\mul / \mub $ shows an asymmetric X-shaped structure with the region of minimum anisotropy offset from the minor axis by about $2\dg$ at high $|b|$. The dispersion ratio is slightly larger than 1.1 along the minor axis and reaches 1.4 at high $|l|$ near the plane of the disc. These features are reproduced well by the model which has slightly lower dispersion ratio around the minor axis.

The correlation maps show a clear quadrupole structure with the magnitude of the correlation at $\approx0.1$. 
The correlation is stronger at positive longitudes which is likely due to the viewing angle of the bar as the model also shows the signature. 
This shows that the bar orbits expand in both $l$ and $b$ while moving out along the bar major axis.
This is consistent with the X-shaped bar but could also be caused by a radially anisotropic bulge so this result in itself is not conclusive evidence for the X-shape. 
However the fiducial model is a very good match to the structure of the observed signal which gives us confidence that this signature is caused by an X-shaped bulge similar to the model. 
In addition, the difference between correlation amplitude between positive and negative longitude rules out a dominant spherical component as this would produce a symmetrical signature.

All of the results of the integrated kinematic moments are consistent with the picture of the bulge predominantly being an inclined bar, rotating clockwise viewed from the north galactic pole, with the near side at positive longitude. The fiducial bar model is a very good match to all of the presented kinematic moments which gives us confidence that the model can provide a quantitative understanding of the structure and kinematics of the bulge.


\subsection{Comparison to Earlier Work}
Previous studies of MW proper motions have been limited to small numbers of fields. Due to the difficulty of obtaining quasars to anchor the reference frame these studies have dealt exclusively with relative proper motions.
In this section we compare VIRAC to two previous studies, \citetalias{kozlowski_2006} and \citetalias{rattenbury_2007}. 
These studies have a relatively large number of fields, 35 HST fields for \citetalias{kozlowski_2006} and 45 OGLE fields for \citetalias{rattenbury_2007}, so on-sky trends are visible. 
Both of these studies have different selection functions from VIRAC and so here we mainly compare the average trends in the data with less focus on the absolute values. 
We do not consider other previous works because in some cases they discuss only results for a single field. 
Comparing kinematics for single fields is less informative due to the effects of the selection functions and other systematics, 
Figure \ref{fig:compare2otherData} shows the comparison of the dispersions, dispersion ratio and correlation measurements from VIRAC with those of \citetalias{kozlowski_2006} and \citetalias{rattenbury_2007}.

We see excellent agreement between the VIRAC data and the \citetalias{rattenbury_2007} measurements in all 4 kinematic moments. 
The dispersion trends are clearly consistent; both VIRAC and \citetalias{rattenbury_2007} dispersion measurements increase towards the MW plane.
The lobe structures caused by the superposition of barred bulge and disc are also reproduced in both the VIRAC data and \citetalias{rattenbury_2007} with the dispersion at high positive $l$ larger than at high negative $l$ for both dispersions. 
The dispersion ratios also match nicely with the lowest ratio found along the minor axis and then increasing for larger $|l|$ sub-tiles. The correlation maps are also in excellent agreement with a clear quadrupole signature visible in both VIRAC and \citetalias{rattenbury_2007}.

The agreement between VIRAC and \citetalias{kozlowski_2006} is less compelling. This is likely due to the larger spread of measurements in adjacent sub-tiles. In the dispersion maps we still see the general increase in dispersion towards the galactic plane, however the trend is far less smooth for the \citetalias{kozlowski_2006} data than for the VIRAC or \citetalias{rattenbury_2007} data. There also appears to be a slight offset in the absolute values although this is expected since VIRAC does not replicate the selection function of \citetalias{kozlowski_2006}.
For the dispersion ratio we observe a similar overall trend; the dispersion ratio increases moving away from the minor axis. This is likely due to the X-shape. There is a single outlying point in the dispertion ratio map at $\sim$($5\dg$,$-4\dg$) that has a ratio  $\approx$0.3 greater than the immediately adjacent sub-tile. This outlier is caused by a high $\dpmb$ measurement. The correlations are in good agreement between the two datasets although the \citetalias{kozlowski_2006} sample only probes the $(+l,-b)$ quadrant.

%% file: sections/e_correlationRG.tex
\begin{figure*}
	\subfigure{\includegraphics[width=\textwidth]{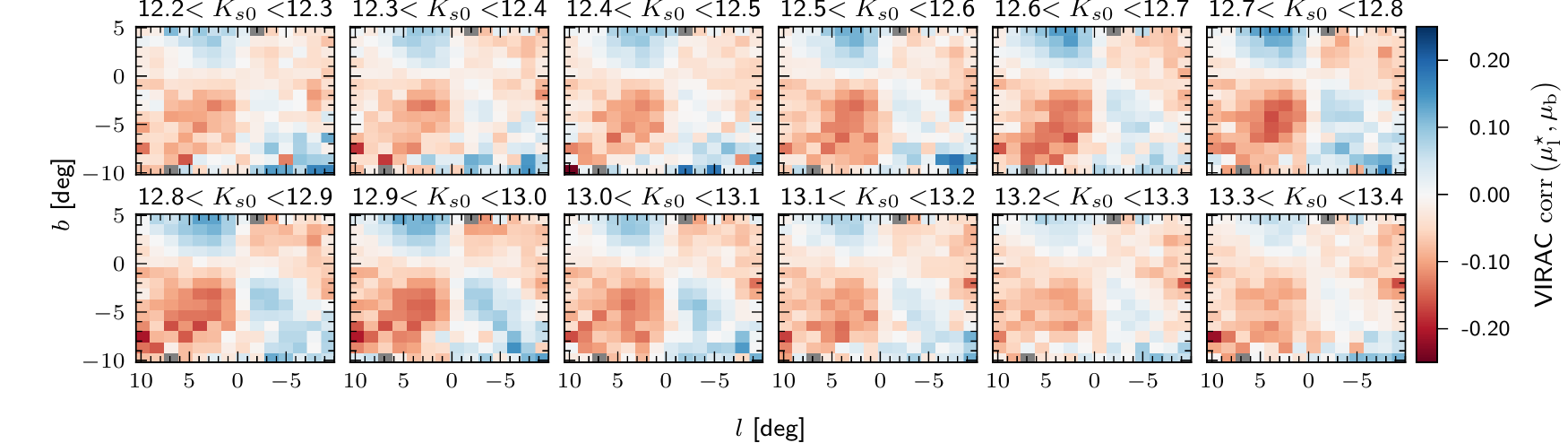}}
	\\[0cm]
	\subfigure{\includegraphics[width=\textwidth]{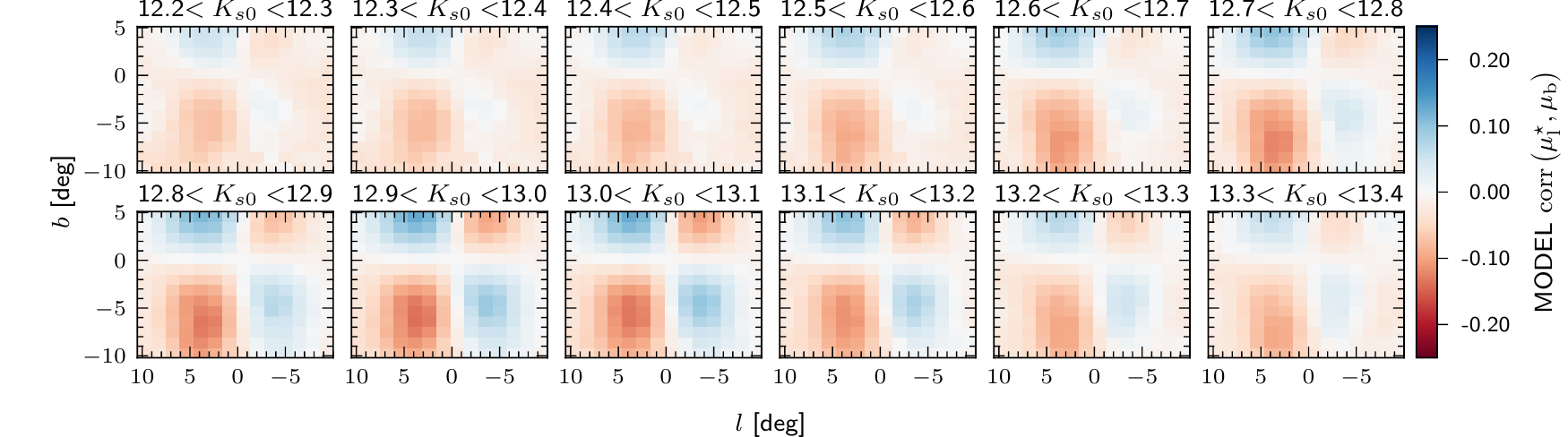}}
    \caption{Correlation of $\mul$ and $\mub$ for the VIRAC data (upper) and the fiducial barred model (lower) in spatial fields on the sky and split into magnitude bins of width $\Delta K_{s0}=0.1$mag. We see the same quadrupole structure in all magnitude bins. In both the data and the model the correlation signal is stronger in the magnitude range 12.5$<K_s<$13.1 mag which corresponds to the magnitude range of the inner bulge RC population. }
    \label{fig:correlation_ksliced}
\end{figure*}

\subsection{Correlation in Magnitude Slices}

In this section we decompose the integrated RGB correlation map into magnitude bins of width $\Delta K_{s0}=0.1$ mag, see figure \ref{fig:correlation_ksliced}. 
As in the integrated map, the magnitude resolved correlation maps all show a distinct quadrupole structure as well as a disparity between the strength of the correlation at positive and negative longitude. The magnitude binning also reveals that the brightest and faintest stars have less correlated proper motions than stars in the magnitude range $12.5<K_{s0}<13.1$ mag which corresponds to the inner-bulge RC stellar population. 
As RC stars have a narrow LF their magnitude can be used as a rough proxy for distance. The rise and fall of the correlation therefore demonstrates that a fraction of RC stars in the
inner bulge ($\pm0.3$ mag $\sim\pm1.2$ kpc along the LOS) have correlated proper motions. This signature is very similar in the analogous plots for the fiducial barred bulge model in figure \ref{fig:correlation_ksliced}. There is no evidence in the VIRAC data that the correlated RC fraction decreases towards the Galactic centre, as would be expected if a more axisymmetric classical bulge component dominated the central parts of the
bulge. In the RGB population, underneath the RC, the correlation is spread out in magnitude because of the exponential nature of the RGB; this plausibly explains the baseline correlation seen at all magnitudes in figure \ref{fig:correlation_ksliced}.

%% file: sections/f_KinematicsGetRc.tex
RC stars are valuable tracers to extract distance resolved information from the VIRAC data. 
They are numerous and, due to their narrow range of absolute magnitudes, their apparent magnitudes are a good proxy for their distance. 
From the LF the combination of RC, RGBB and AGBB is readily obtained with the fraction of contaminating stars relative to the RC $\sim 24\%$ consistent with RGBB measurements from \citet{nataf_2011}, see also section \ref{sec:m2m}.
It is possible to obtain an estimate for just the RC from the RC\&B using a deconvolution procedure as used in \citetalias{wegg_2013} however we do not do this here.

\subsection{Structure of the Red Giant Branch Continuum} \label{subsec:structureRGBC}

The RGBC absolute LF, as discussed in section \ref{subsec:particle2stellarDist}, is well described by an exponential function. 
We assume that the stellar population is uniform across the entire MW bulge distance distribution and therefore there exists a uniform absolute magnitude LF for the RGBC,
\begin{equation}
  \Lagr\left(\mk\right) \, \propto \, e^{\,\beta \mk},
\end{equation}\label{eqn:absLF}
where $\beta$ is the exponential scale factor, see equation \ref{eqn:rgbc_lf}. 

We now demonstrate that the proper motion distribution of the RGBC is constant at all magnitudes. This will allow us to measure the proper motion distribution of the faint RGBC, where there is no contribution from the RC\&B, and subtract it at all magnitudes. The result is the proper motion distribution as a function of RC standard candle magnitude with only a small contamination from RGBB and AGBB stars.

Consider two groups of stars at distance moduli $\mu_1$ and $\mu_2$ with separation $\Delta\mu=\mu_2-\mu_1$. These groups generate two magnitude distributions $\Lagr_1 \propto 10^{\beta \mu_1} $ and $\Lagr_2 \propto 10^{\beta \mu_2} $ respectively. 
$\Lagr_2$ can be rewritten as,
\begin{equation}
    \Lagr_2 \propto 10^{\beta (\Delta\mu + \mu_1)} \propto 10^{\beta\Delta\mu}10^{\beta\mu_1},
\end{equation}
meaning both groups of stars produce the same magnitude distribution but with a relative scaling that depends upon the distance separation and the density ratio at each distance modulus. 
Generalising this to the bulge distance distribution; each distance generates an exponential luminosity function that contributes the same relative fraction of stars to each magnitude interval. 
This is also true for the velocity distributions from the various distances and so we expect the velocity distribution of the RGBC to be the same at all magnitudes.

\begin{figure}
	\includegraphics[width=\columnwidth]{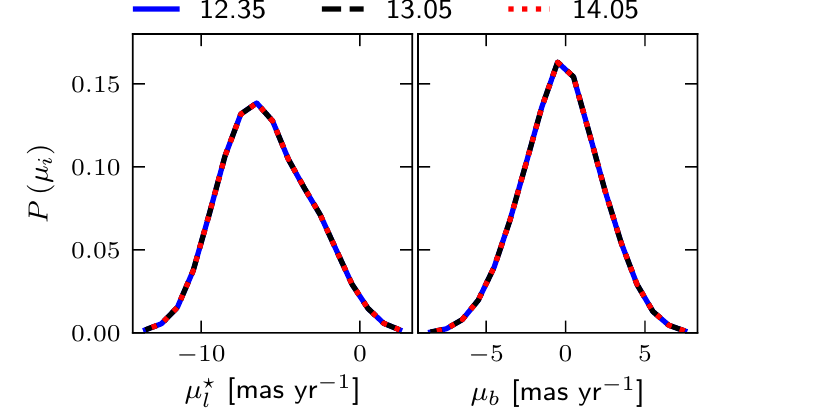}
    \caption{ 
    Histograms of the RGBC proper motion distributions from the model at three magnitude intervals, along a single LOS, considering all model disk and bulge particles.
    The histograms are individually normalised and clearly show that the three profiles lie directly on top of each other. This is the case for all magnitude intervals we are considering. The proper motion distribution at each magnitude has the same structure but the overall normalisation changes allowing the distribution at faint magnitudes without RC\&B contamination to be used at brighter magnitudes.
    }
    \label{fig:bkg_hypothesis}
\end{figure}

To test this further we construct the RGBC ($\mu_{l,b}$,$K_{s0}$) distributions for a single LOS using the model and the RGBC absolute LF constructed in section \ref{subsec:particle2stellarDist}. We then normalise the distributions for each magnitude interval individually and the distributions for three magnitudes are shown in figure \ref{fig:bkg_hypothesis}. This shows that the RGBC proper motion distributions are magnitude independent. The distribution at faint magnitudes, $14.1<K_{s0}<14.3$ mag, where there is no contamination from the RC\&B, can be used to remove the RGBC at brighter magnitudes where the RC\&B contributes significantly.

\subsection{Extracting the Kinematics of the RC\&B}\label{subsec:extractingVIRACkinematics}

\begin{figure}
	\includegraphics[width=\columnwidth]{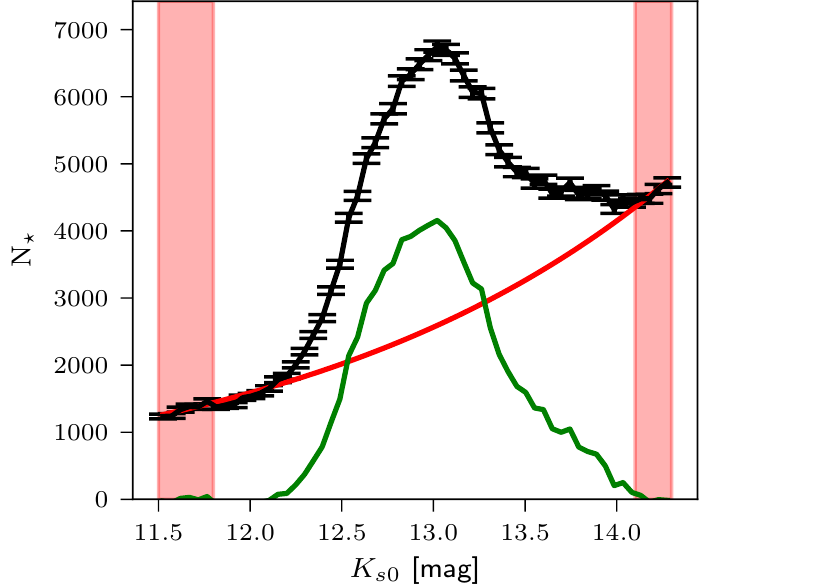}
    \caption{ (b278 (1.$^\circ$, -4.2$^\circ$)) This plot shows the fit to the RGBC for the example tile in the VIRAC data. We use two magnitude intervals, 11.5$<K_s<$11.8 and 14.1$<K_s<$14.3 mag, shown as the red regions for the fitting. Subtracting the fit, red line, from the tile LF, shown in black, gives the LF of the RC\&B.}
    \label{fig:bkg_fitting_VVV}
\end{figure}

We have just shown that the proper motion distribution of the RGBC at faint magnitudes, where it can be directly measured, is an excellent approximation of the proper motion distribution at brighter magnitudes where it overlaps with the RC\&B. 
We use this to subtract the RGBCs contribution to the VIRAC magnitude - proper motion distributions. 
The first step is to fit the RGBC LF marginalised over the proper motion axis. 
This provides the fraction of RGBC stars in each magnitude interval relative to the number of RC\&B stars. 
We fit a straight line to $\log(N_{ \mathrm{RGBC} })$,
\begin{equation}
  \log(N_{ \mathrm{RGBC} }) = A + B \left( K_{s0} - K_{s0, \mathrm{RC} } \right),
\end{equation}
where $A$ and $B$ are the constants to be fitted and $K_{s, \mathrm{RC} }=13.0$ mag is the approximate apparent magnitude of the RC. 
When fitting, we use the statistical uncertainties from the Poisson error of the counts in each bin.
The LF is fitted within two magnitude regions on either side of the clump; $11.5<K_{s0}<11.8$ and $14.1<K_{s0}<14.3$ mag. 
The bright region is brighter than the start of the RC over density but is not yet affected by the saturation limit of the VVV survey.
The faint region is selected to be fainter than the end of the RGBB but as bright as possible to avoid uncertainties due to increasing incompleteness at faint magnitudes. 
The fit for the example tile is shown in figure \ref{fig:bkg_fitting_VVV}. Included are the two fitting regions in red and the RC\&B LF in green following the subtraction of the fitted RGBC.

\begin{figure*}
	\includegraphics[width=\textwidth]{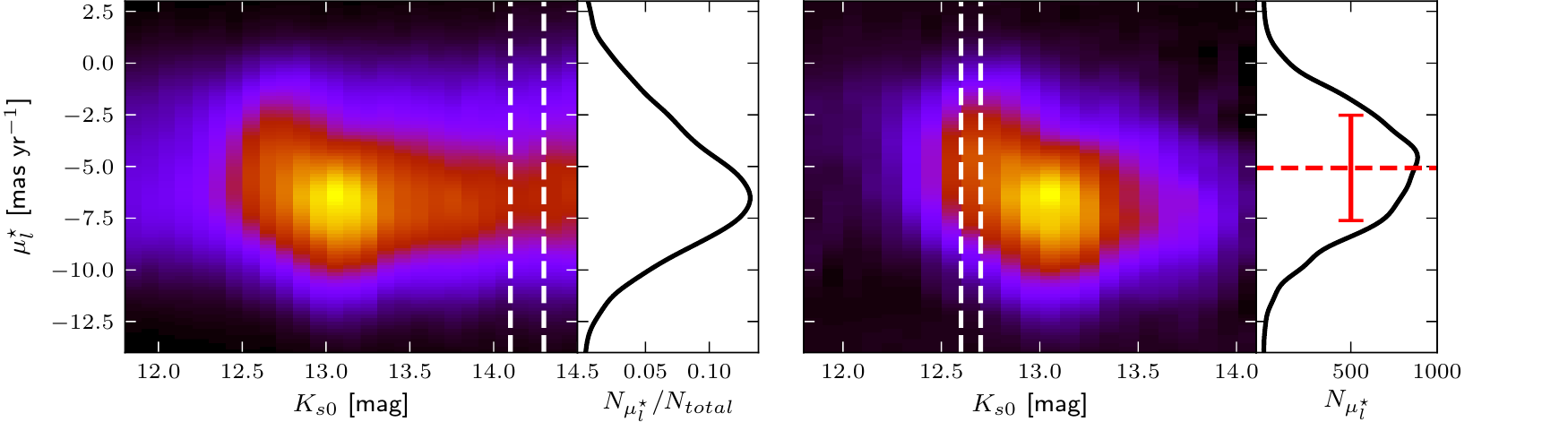}
    \caption{ 
    (b278 (1.$^\circ$, -4.2$^\circ$)) Process for extracting the kinematics as a function of magnitude for the RC\&B from the total RGB ($K_{s0}$,$\mul$) distribution. 
    Left plot: The kernel density smoothed RGB distribution (left panel) with white lines highlighting the magnitude interval used for constructing the proper motion distribution of the RGBC, (right panel). This RGBC distribution is subtracted at each magnitude normalised according to the RGBC fit. Right plot: The ($K_{s0}$,$\mul$) distribution (left panel) for the RC\&B following the subtraction of the RGBC. The vertical white lines highlight a magnitude bin for which the kinematic measurements are shown (right panel). The horizontal dashed line shows the mean, and the error bar shows the dispersion.
    }
    \label{fig:faint_bkg_subtraction}
\end{figure*}

The second step to extract the RC\&B velocity distribution is to remove the RGBC velocity distribution. 
This process is summarised in figure \ref{fig:faint_bkg_subtraction}. 
We construct the RGBC velocity distribution using a kernel density estimation procedure.
For consistency we compute the RGBC proper motion profile using the same faint magnitude interval used for the RGBC fitting.
The background is scaled to have the correct normalisation for each magnitude interval according to the exponential fit.
The total proper motion profile for each magnitude interval is then constructed using the same kernel density estimation procedure. 
We use a rejection sampling approach to reconstruct the RC\&B proper motion distribution with discrete samples. We sample two random numbers:
\begin{inparaenum}
\item The first in the full range of proper motions covered by the two proper motion distributions, total distribution and the scaled RGBC distribution, in the magnitude interval.
\item The second between zero and the maximum value of the two kernel density smoothed curves.
\end{inparaenum}
Only points that lie between the two distributions, in the velocity range where the two distributions are statistically distinct, are kept, as only these points trace the RC\&B distribution. We sample the same number of points as the exponential fit indicates there are in the RC\&B component. This is to reconstruct the distribution with the correct level of accuracy. 
For this sample of points we compute the mean and dispersions analytically.
We repeat this sampling in a Monte Carlo procedure to obtain 100 realisations of the mean and dispersion measurements and use these to characterise the uncertainty upon the measurements.

This approach ignores the variable broadening as a function of magnitude caused by measurement uncertainties. To test this we extracted the magnitude-proper motion data from the model for a variety of representative tiles and convolved the values with the median VIRAC uncertainties.
The convolution increases the dispersion by $~0.06$ $\masyr$ at $K_{s0}=11.8$ mag and $\sim0.16\pm0.05$ $\masyr$ at $K_{s0}=13.6$ mag. The broadening at fainter magnitudes is more sensitive to the spatial location of the tile.
The model provides discrete samples of the RC\&B kinematic distribution as a function of magnitude and so we calculated the convolved mean proper motions and dispersions analytically.
We then applied the same analysis as described for the data to the complete convolved distribution drawn from the model, disregarding the known separation between RC\&B and RGBC.
Comparing the analytically calculated kinematics with the data-method measurements we find a systematic uncertainty in the recovered values of $\lesssim$0.1 $\masyr$ for dispersion and significantly less for the mean. This systematic can be positive or negative for a given tile but is consistent at all magnitude intervals along the LOS.

\begin{figure*}
  \centering
  \subfigure{\includegraphics[width=\textwidth]{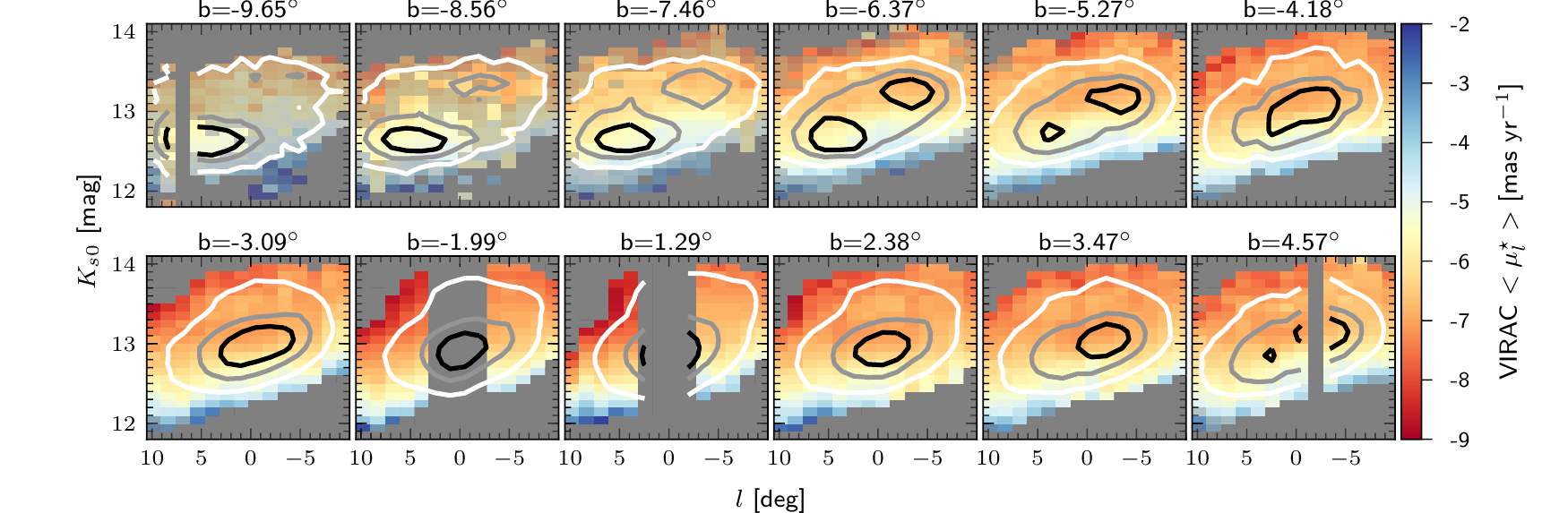}}
  \subfigure{\includegraphics[width=\textwidth]{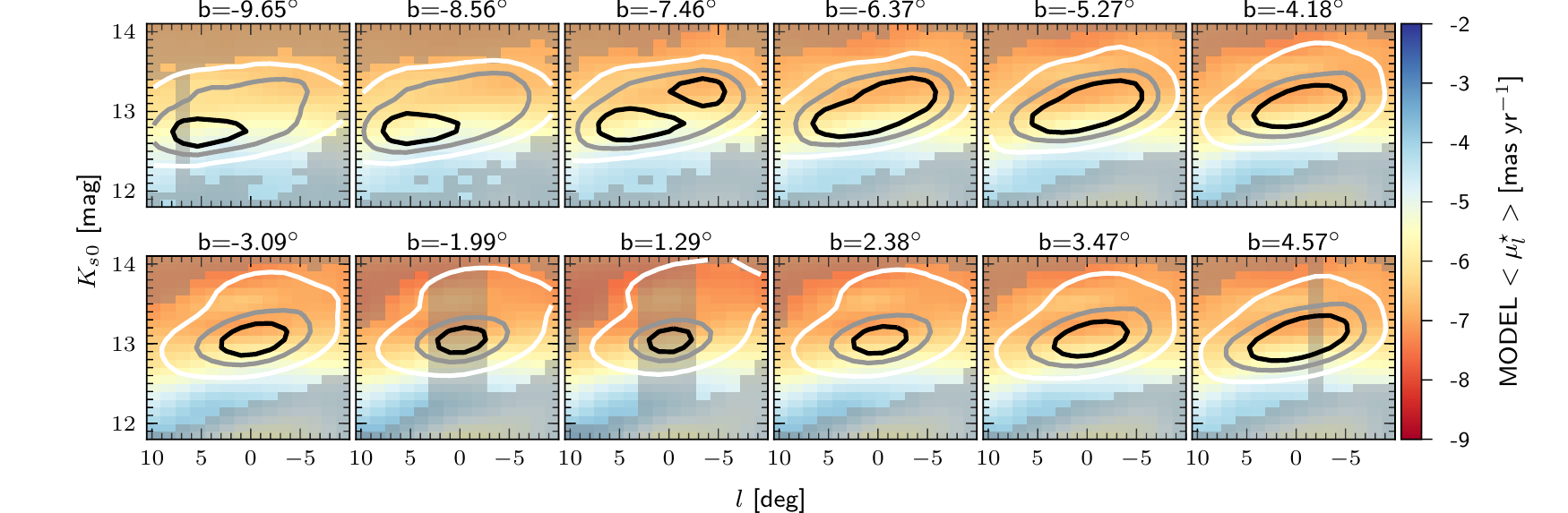}}
  \caption{ Top panels: $\mpml$ maps of the RC\&B stars in latitude slices as a function of magnitude for the VIRAC data. The contours correspond to the stellar number count of the RC\&B stars. Focusing on the top row in particular where we observe a split RC\&B we see that the two density peaks have $\dmpml$ $\approx$ 1 $\masyr$.
  Lower panels: Equivalent plots for the fiducial bar model from \citetalias{portail_2017} which matches the mean transverse motion and the gradients in the data very well. The grey areas in the VIRAC plots are masked based on our measurement errors and are shaded in the model plots to guide the eye.}
  \label{fig:rcb_meanPM_L_vvv}
\end{figure*}

%% file: sections/g_kinematics_Bsliced.tex
\begin{figure}
	\includegraphics[width=\columnwidth]{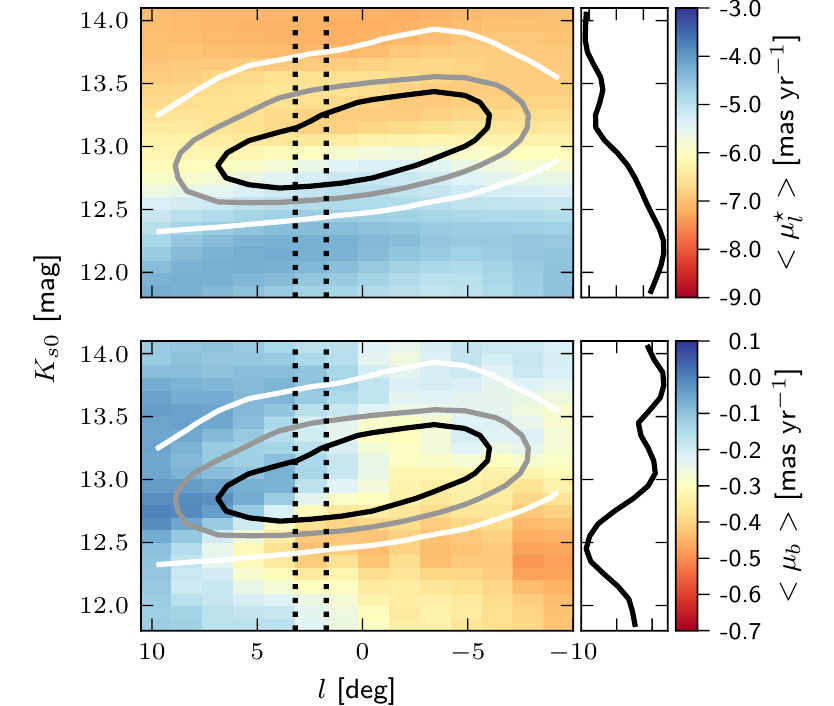}
    \caption{Zoom in of $b=-6.37\dg$ slice for the model $\mpml$ (top panel) and $\mpmb$ (bottom panel). 
    The panels to the right show the profile for the tile highlighted by the dotted lines. The profiles show a clear series of kinks rather than a smoothly varying structure which are consequences of the pattern rotation, streaming motions along the bar and the presence of multiple stellar evolutionary stages.  
    The contours show the RC\&B star counts. }
    \label{fig:zoom_in_4_streaming}
\end{figure}

\begin{figure*}
  \centering
  \subfigure{\includegraphics[width=\textwidth]{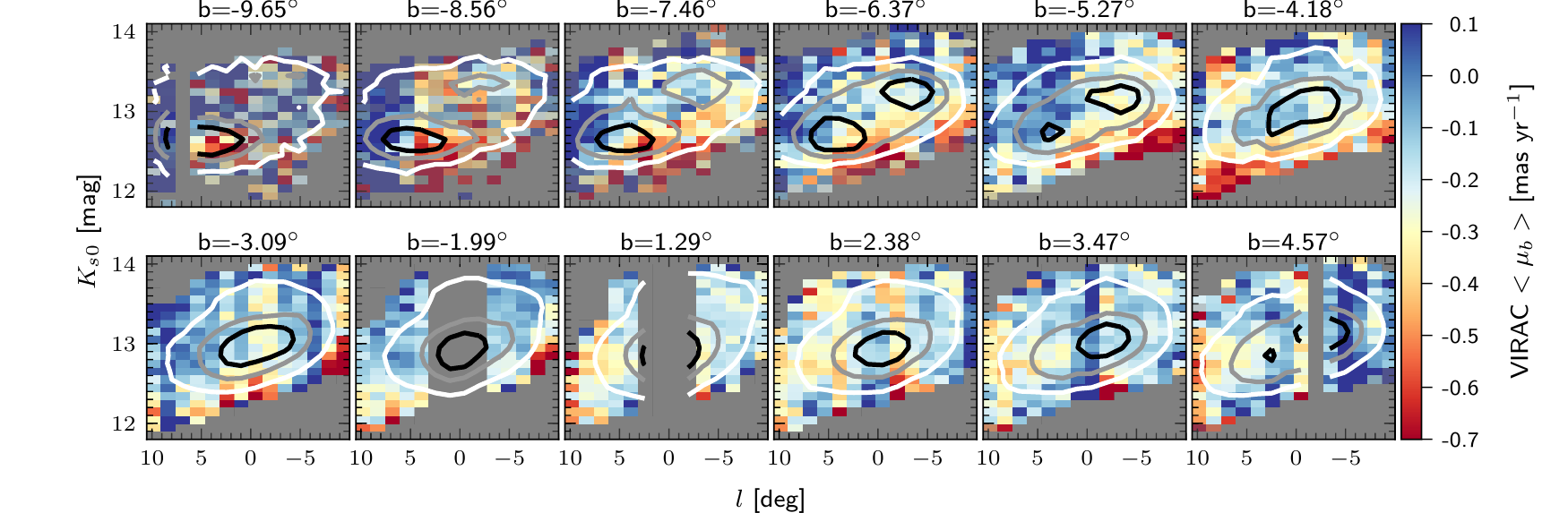}}
  \subfigure{\includegraphics[width=\textwidth]{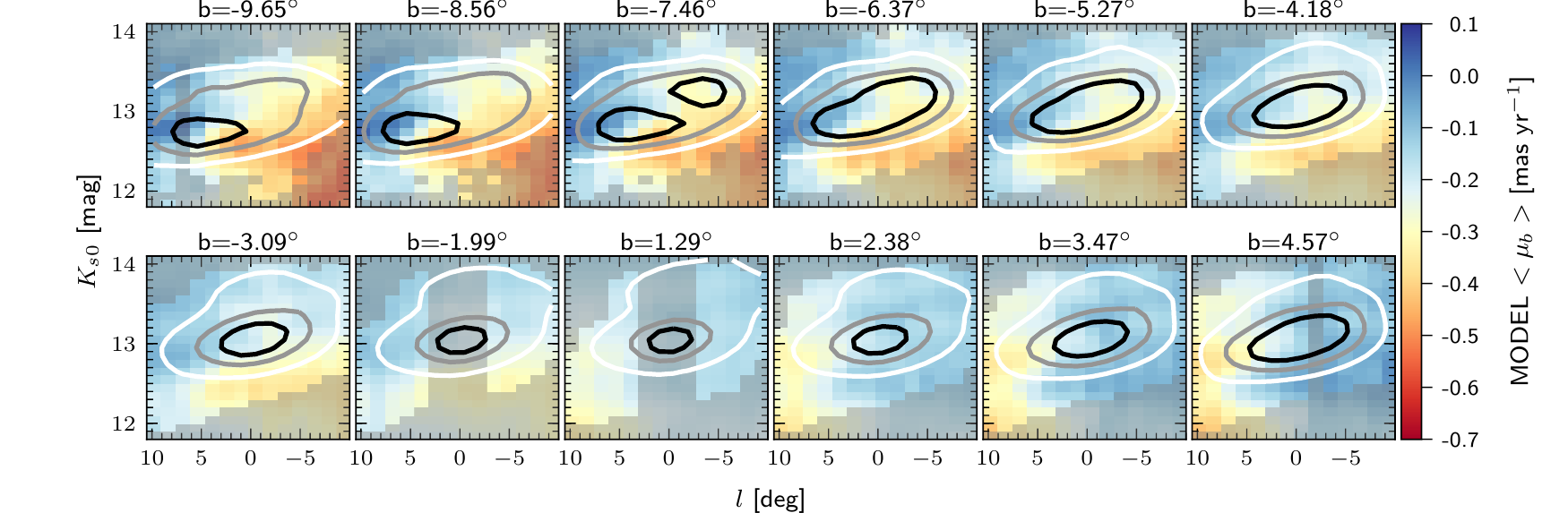}}
  \caption{$\mpmb$ maps of the RC\&B stars as a function of magnitude for the VIRAC data with the format of the plots identical as in figure \ref{fig:rcb_meanPM_L_vvv}. The model reproduces the transition between more positive to more negative proper motion aligned with the bar axis shown by the star count contours. This pattern reflects the streaming motion within the bar and the bar pattern rotation. }
  \label{fig:rcb_meanPM_B_vvv}
\end{figure*}

\begin{figure*}
  \centering
  \subfigure{\includegraphics[width=\textwidth]{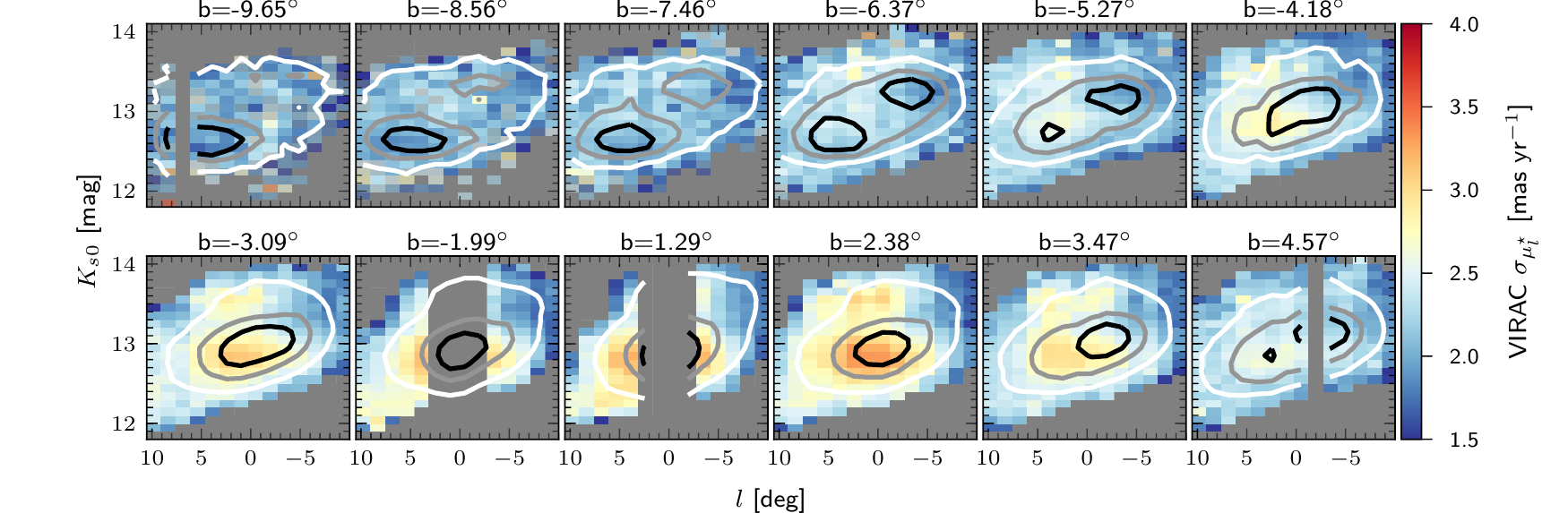}}
  \subfigure{\includegraphics[width=\textwidth]{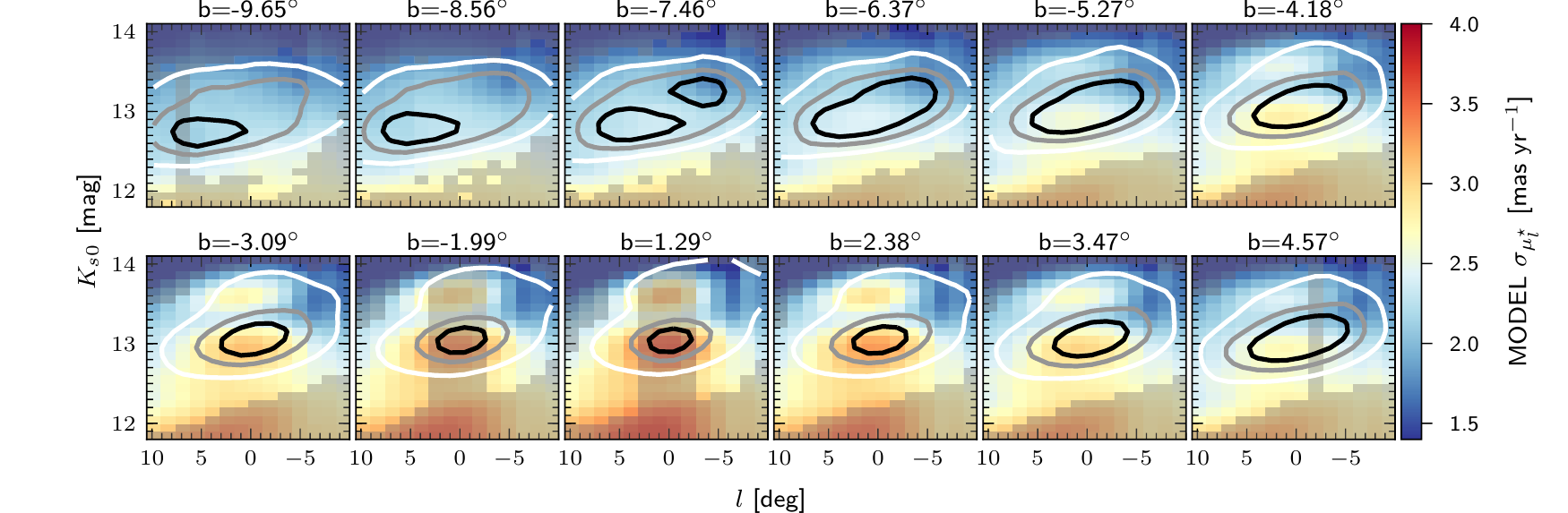}}
  \caption{$\dpml$ maps of the RC\&B stars as a function of magnitude for the VIRAC data with the format of the plots identical as in figure \ref{fig:rcb_meanPM_L_vvv}. The model nicely reproduces all of the features such as the central dispersion peak due to the RC stars in the galactic centre, the secondary peak corresponding to the fainter RGBB stars also in the galactic centre, and the increased dispersion gradient in the bar starting at $|b|\approx 6^\circ$ which is caused by the intrinsic proper motions of stars in the bar beginning to dominate. }
  \label{fig:rcb_dispPM_L_vvv}
\end{figure*}

\begin{figure*}
  \centering
  \subfigure{\includegraphics[width=\textwidth]{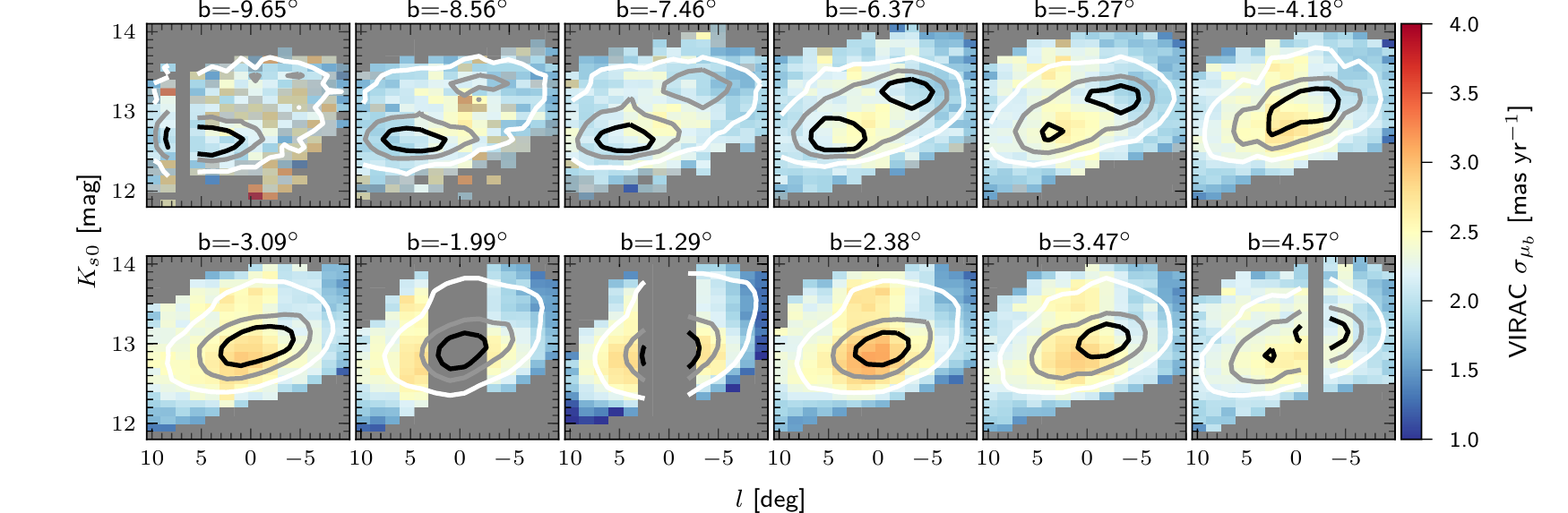}}
  \subfigure{\includegraphics[width=\textwidth]{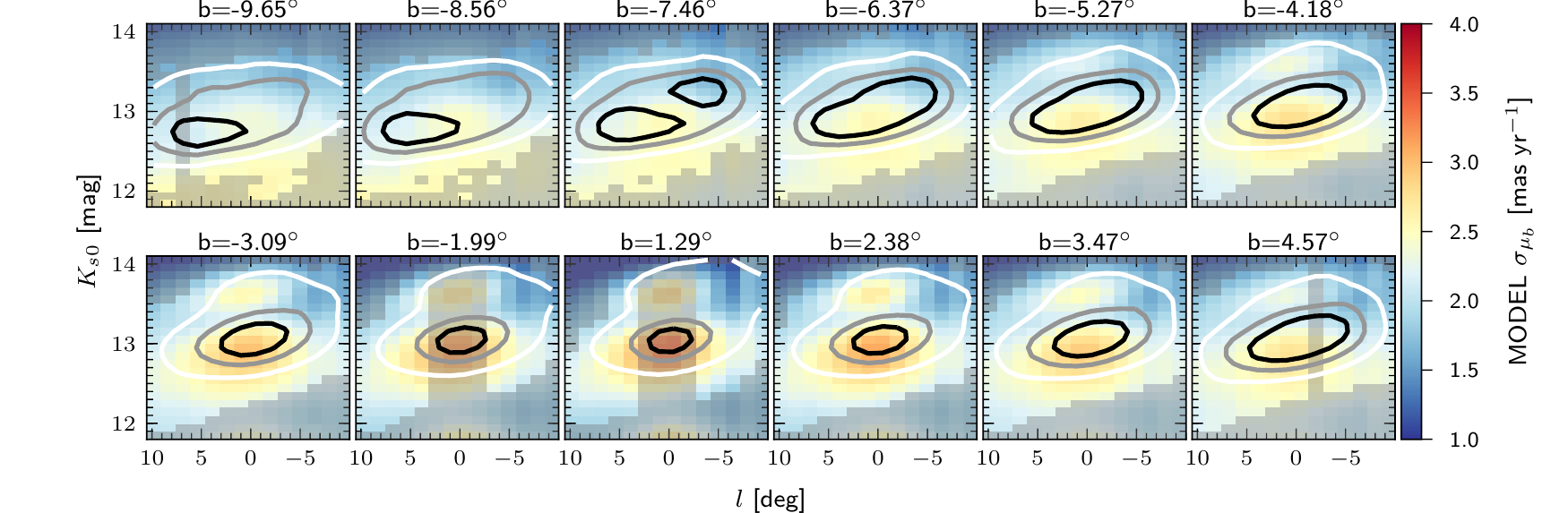}}
  \caption{$\dpmb$ maps of the RC\&B stars as a function of magnitude for the VIRAC data with the format of the plots identical as in figure \ref{fig:rcb_meanPM_L_vvv}. As in figure \ref{fig:rcb_dispPM_L_vvv} the model produces an excellent match to the structures seen in the data. }
  \label{fig:rcb_dispPM_B_vvv}
\end{figure*}
\subsection{Latitude Slices}



The luminosity function along the minor axis for high latitude tiles in the bulge exhibits a double peaked distribution which is believed to be due to an X-shaped boxy/peanut bulge. The acute viewing angle of the bar causes lines of sight near the minor axis at high latitude to intersect the near arm first and subsequently the faint arm of the X-shape. As discussed in the Introduction, this scenario is supported by various evidence from observations and N-body simulations, but alternative scenarios based on multiple stellar populations along the line-of-sight have also been suggested. In this section we present proper motion kinematics of RC\&B stars as a function of magnitude which provide an independent test of these scenarios.

Figures \ref{fig:rcb_meanPM_L_vvv} to \ref{fig:rcb_dispPM_B_vvv} show the number density, mean proper motions and proper motion dispersions of RC\&B stars in latitude slices as a function of magnitude for both VIRAC and the fiducial dynamical model from P17.

In section \ref{sec:getRCB} we described the rejection sampling approach to measure the proper motion mean and dispersion. We apply an opaque mask to bins in which the RC\&B contributes less than 10\% of the stars according to the RGBC fit to ensure that the results are reliable. We apply a secondary transparent mask to all regions where the Monte Carlo resampling measurement uncertainty is greater than 0.1 $\masyr$ to guide the eye as to where the results are most secure.
As mentioned in section \ref{sec:getRCB} there is also a systematic uncertainty of at maximum 0.1 $\masyr$ in the dispersion measurements and smaller for the mean measurements which is caused by the magnitude dependent broadening of the proper motion distributions.

The fiducial model has been fitted to star count data and radial velocity
data for the bulge and long bar as described in Section \ref{sec:m2m}, but no VIRAC proper motion data was used. It nonetheless provides excellent predictions for the observed PM data, and can therefore be used to understand the signatures present in the VIRAC maps.

\subsubsection{Number Density}\label{subsec:RCdensity}

The star counts of RC\&B stars are shown with the grey contours in figure \ref{fig:rcb_meanPM_L_vvv}. 
N ear the minor axis at $|b|>4.^\circ$ the contours show a bi-modal star count distribution while at $|b|>6.^\circ$ they show clear evidence of double peaked luminosity functions. 
These results are both consistent with \citet{saito_2011} and \citetalias{wegg_2013}, who studied the distribution of RC stars using VVV, and with previous studies \citep{mcwilliam_2010,nataf_2010}. As expected they are consistent with the structure of a boxy/peanut bulge with the near end at positive longitude. The model, which is known to host an X-shaped structure, nicely replicates the extension of the final density contour towards fainter magnitudes which is caused by the presence of the RGBB stars.

\subsubsection{Mean Longitudinal Proper Motion}

The VIRAC $\mpml$ of the RC\&B as a function of tile and magnitude is shown in the upper plot of figure \ref{fig:rcb_meanPM_L_vvv}.
The overall proper motion of the galactic centre is consistent with the solar reflex motion $\mul=-6.38$ $\masyr$ \hbox{\citep{reid_2004}}.
We see that at all latitudes the brighter stars have a less negative proper motion than the fainter stars and the observed gradient is well reproduced by the model. 

A zoom in of the $b=-6.37\dg$ slice for the model $\mpml$ is shown in the top panel of figure \ref{fig:zoom_in_4_streaming}.
The overall bright to faint $\mpml$ gradient shows the mean rotation of stars as a function of distance which is lower than for circular orbits in a disk. The barred structure causes a longitudinally asymmetric pattern different from expected for a circular rotation field. These features are sensitive to the pattern rotation and to streaming motions in the bar.
The effect of streaming can be seen at $|l|\lesssim4\dg$. Considering $\mpml$ there is a smooth but rapid transition from more positive to more negative $\mul$ between $12.2<K_{s0}<13.2$ mag where the mean is dominated by the RC. 
This is followed by a kink at $K_{s0}\sim13.5$ where the RGBB stars in the near side region of high bulge density cause a kink towards more positive mean proper motion. 
Ther initial transition is much stronger in the tiles near the minor axis, $|l|\lesssim4\dg$, and the kinks are only observed in this region. The kinks being longitude dependant makes this unlikely to be a purely stellar population effect. We expect the greatest streaming velocities near the minor axis and so it is likely that a combination of stellar type and streaming is causing these effects.
This kink in the proper motion profiles as a function of $K_{s0}$ can also be seen in the VIRAC data in figure \ref{fig:rcb_meanPM_L_vvv}.
At bright magnitudes $\mpml$ becomes more negative again due to AGBB stars in the high density bulge region which have more negative proper motions than the closer RC and RGBB stars.

At higher latitudes that exhibit a double peaked density distribution the misalignment of the proper motion transition causes the brighter peak to have mean proper motion $\approx1$ $\masyr$ more positive than the fainter peaks. 
This demonstrates that the bright peak in the split RC has significantly distinct proper motion kinematics from the faint peak. 
The faint and bright RC division can therefore not have a purely stellar population origin.
Instead, the observed effects are well reproduced by the X-shaped bar model, shown in the lower plots. 
Since the barred potential and the orbits in it are largely fixed by the fitted data, and both RC peaks are visited by similar orbits \citep{portail_2015b}, it is hard to see how the barred model could support the split RC peaks through different stellar populations.

\subsubsection{Mean Latitudinal Proper Motion}
The VIRAC $\mpmb$ of the RC\&B as a function of tile and magnitude is shown in the upper plot of figure \ref{fig:rcb_meanPM_B_vvv}. 
The $\mpmb$ appear noisier compared to $\mpml$ because while both maps are subject to systematic errors of $\approx$0.1 $\masyr$, $\mpmb$ covers a smaller range of values.
The systematics are a combination of the relative to absolute correction, see section \ref{sec:vvvpm}, and the effect of variable broadening on our RC\&B extraction approach, see section \ref{sec:getRCB}. 
The reflex motion due to the sun's vertical motion is $\approx-0.2$ $\masyr$ for $V_{z,\odot}=7.25 \, \kms$ \citep{schoenrich_2010} which broadly accounts for the overall offset from zero in the fiducial model shown in the lower plots. 
At latitudes $|b|>4^\circ$ the $\mpmb$ isocontours for both VIRAC and the model highlight a transition that is aligned with the bar axis shown in the star count contours. 
Considering the zoom in of the $b=-6.37\dg$ slice for the model $\mpmb$ shown in the bottom panel of figure \ref{fig:zoom_in_4_streaming}, the near side of the bar, along the $l=2.5\dg$ LOS, shows strong negative $\mpmb$ while the far side shows more positive $\mpmb$. If only pattern rotation were contributing we would expect a smoothly declining trend as the apparent proper motion decreases for stars at greater distance. 
At this latitude the strong variation in $\mpmb$ is plausibly explained by streaming motions. Specifically, streaming motions in the near side towards the Sun induce an apparent negative $\mub$ while streaming motion away from the sun on the far side induce an apparent positive $\mub$.
We see further evidence in figure \ref{fig:zoom_in_4_streaming} with a spur of more negative $\mpmb$ that is located at $|l|\lesssim3.0\dg$ and $K_{s0}\sim13.3$ mag. 
This feature is caused by RGBB stars in the near half of the bar which are streaming towards the Sun and so present a negative $\mpmb$. 

These $\mpmb$ motions in the VIRAC data in figure \ref{fig:rcb_meanPM_B_vvv} are therefore due to a superposition of streaming velocities in the bar frame along the LOS as well as the bar pattern rotation.
We see similar features of streaming motions in the model, including at latitudes closer to the plane where they are not visible in the VIRAC data for our magnitude range. 

\subsubsection{Longitudinal Proper Motion Dispersion}

$\dpml$ as a function of tile and magnitude for the RC\&B is shown in the upper plot of figure \ref{fig:rcb_dispPM_L_vvv} and corresponding plots for the fiducial model are shown below.  
We see a clear centrally concentrated dispersion peak for tiles close to the plane. This dispersion peak is reproduced by the model where it is caused by the depth of the central potential as opposed to being a separate bulge component.
For latitudes in the range $3<|b|<6\dg$ there is a clear gradient in the dispersion between the near side of the bar and its far side which is at lower dispersion. 
This is reproduced by the model and is because, while the RC\&B stars on both sides have symmetric intrinsic dispersion, the greater distance for the far side of the bar makes the dispersion appear smaller. 
The dispersion gradient becomes less pronounced beyond $|b|>6^\circ$ for both VIRAC and the model. 
For latitudes $|b|<4\dg$ there is a secondary peak of high dispersion $\sim0.8$ mag fainter than the central peak at $K_{s0}=12.7$ mag. 
This is caused by the RGBB stars near the galactic centre.

\subsubsection{Latitudinal Proper Motion Dispersion}

$\dpmb$ as a function of tile and magnitude for the RC\&B is shown in the upper panels of figure \ref{fig:rcb_dispPM_B_vvv}. The latitudinal dispersions show structures very similar to those in the longitudinal dispersion maps. 
We see a concentrated central peak due to RC stars in the deep potential well near the galactic centre and a fainter second peak which is caused by the RGBB stars. These features are well reproduced by the model which is shown in the lower panels. 
There is a clear gradient between the two ends of the bar for latitudes $|b|>4\dg$ with more distant stars having smaller proper motion for the same intrinsic dispersion.
A notable difference to the longitudinal maps is the shallower gradient in the dispersion between brighter and fainter magnitudes. This is likely due to the foreground bar component having a small vertical dispersion in comparison to that of the X-shaped boxy/peanut bulge.

%% file: sections/h_kinematics_Ksliced.tex
\begin{figure*}
    \centering
    \subfigure{
    \includegraphics[width=\textwidth]{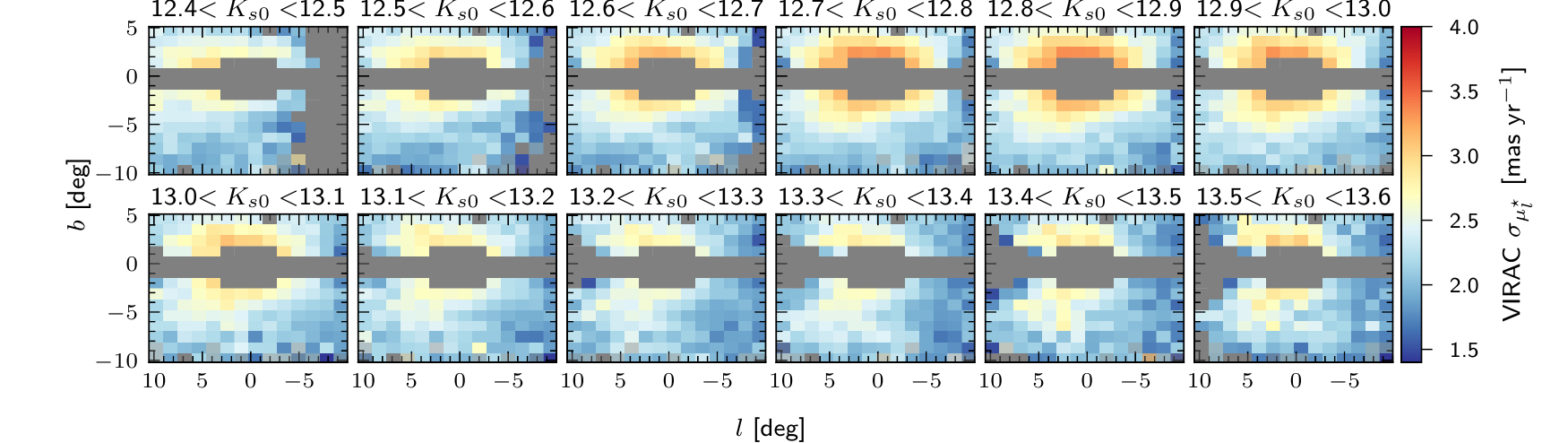}
    }
    \subfigure{
    \includegraphics[width=\textwidth]{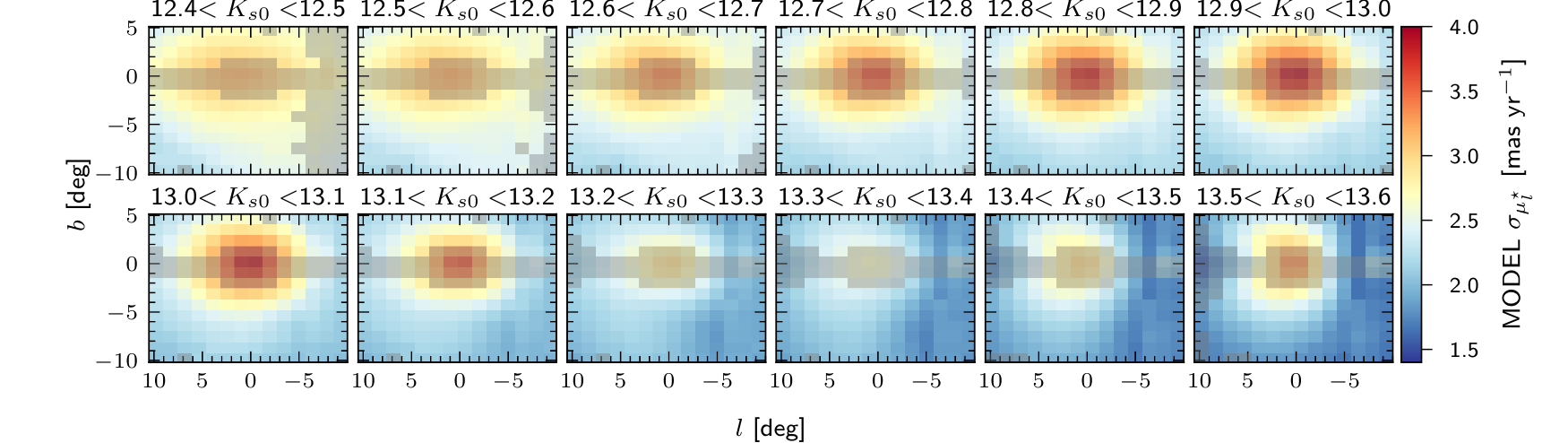}
    }
    \caption{
    This shows $\dpml$ as on-sky maps as a function of magnitude for the data (top panels) and for the barred particle model (bottom panels).
    The difference at high latitude is due to the treatment of the disc in the model which we do not differentiate from the bulge.
    This map helps us to understand the structures seen in figure \ref{fig:integrated_maps}. They show the arched structure at negative latitude only occurs at fainter magnitudes suggesting that the arc is caused by the low dispersion in the far side of the bar.
    }
    \label{fig:dispMul_kslice}
\end{figure*}
\subsection{Magnitude Slices}
Figure \ref{fig:dispMul_kslice} shows the breakdown of longitudinal dispersion in different magnitude intervals for the data (top panels) and fiducial bar model (bottom panels). At all magnitudes we see a high dispersion peak at the galactic centre which is caused by the deep potential well and stars orbiting aligned to the bar major axis. 
This peak is offset slightly towards positive longitude due to the acute observation angle of the bar. 
The magnitude of this peak is strongest at $K_{s0}\approx12.8$ mag which corresponds to RC stars in the centre. The central peak dispersion decreases until $K_{s0}\approx13.3$ mag at which point the dispersion increases again due to RGBB stars in the galactic centre.
We see excellent agreement with the fiducial bar model which reproduces the two central dispersion peaks. 
The model reproduces the arc of low dispersion at negative longitude which is likely caused by the low dispersion of the far side of the bar. The high dispersion peak at brighter magnitudes is not symmetric about the minor axis with near plane positive longitude regions at higher dispersion than their counterpart at negative longitude. This is likely due to the intrinsic dispersion of the near side of the bar.
This plot is complementary to the integrated map, see figure \ref{fig:integrated_maps}, showing the origin of the dynamically colder region at $|b|>5^\circ$ is not a single feature of the bar but rather a superposition of the kinematics at different magnitude intervals.

%% file: sections/z_KinematicsConclusion.tex
We have combined VIRAC and \textit{Gaia} data to obtain $\sim40\,000\,000$ absolute proper motions in 196 tiles to investigate the $-10<l<10^\circ$, $-10<b<5^\circ$ region of the MW barred bulge.

We apply a colour selection to obtain a clean sample of bulge stars and correct for extinction assuming a single foreground sheet. 
We present integrated on-sky maps for the mean proper motions, the proper motion dispersions, the dispersion ratio and correlation.
As a function of magnitude we present on-sky correlation maps of the RGB, and RC\&B mean proper motions and dispersions. 
We derive combined kinematics of the RC, RGBB and AGBB (RC\&B) as a function of magnitude which is a good proxy for the distance due to the small width of the RC luminosity function. These kinematics are presented in latitude slices with longitudinal dispersion also presented in magnitude slices.
The main scientific results of our analysis are:
\begin{itemize}
    \item The $\mpml$ isocontours in the integrated $\mpml$ map are tilted, due to the streaming motions in the bar. The $\mpmb$ map shows a quadrupole signature caused by the composite effect of the bar pattern rotation and longitudinal streaming motions in the bar.
    \item There is a peak in on-sky integrated proper motion dispersions, with $\sigma_\mu>3.$ $\masyr$, at the galactic centre. This is due to the deep potential well which causes stars following bar orbits to pass rapidly through the centre. The dispersion maps exhibit a lobed structure at negative $l$ where $\sigma_\mu$ is $\sim0.2$ $\masyr$ smaller than at positive $l$ for $b=-5\dg$. 
    \item The dispersion ratio exhibits a clear X shape, slightly asymmetrical about the minor axis due to bar geometry, that has minimum $\dpml/\dpmb \approx1.1$ located at $\sim(2\dg,-7\dg)$, and maximum, $\sim1.4$, near the disk for $|l|\gtrsim6\dg$.
    \item There is a distinct quadrupole signature in the integrated correlations which we interpret as being caused by stars following boxy orbits within the bar. The correlation is stronger at $l>0\dg$ as expected for a bar with the near side at $l>0\dg$.
    \item We see an increase in the correlation of RGB star proper motions at magnitudes corresponding to RC stars near the galactic centre. This demonstrates that a significant fraction of stars in the inner bulge have correlated proper motions. Furthermore, we see no decrease in correlation towards the centre which would be expected if a separate, more axisymmetric, classical bulge component dominated in the central parts of the bulge.
    \item In constant latitude slices VIRAC shows a bi-modal star count distribution with clear evidence for the double peaked RC near the minor axis at high latitudes which is consistent with previous work.
    \item $\mpml$ in slices of RC magnitude shows clear evidence of a proper motion difference of $\sim1\masyr$ between the RC\&B stellar populations in the near and far sides of the bar. 
    This strongly supports the X-shaped scenario in which the different sides of the bar move in different directions relative to the Sun. The split RC cannot be explained purely by a population effect.
    \item The overall gradient is sensitive to the pattern rotation and the tilt of the $\mpml$ isocontours is due to the presence of the bar. The $\mpml$ profile along a LOS is sensitive to streaming motions within the bar.
    \item $\mpmb$ shows a gradient aligned with the bar star count contours. We interpret this as evidence for streaming motion in the bar.
\end{itemize}

In parallel we have used an existing barred dynamical model, from \citetalias{portail_2017}, and replicated the selection function of the VIRAC survey, to compare with the observed kinematic maps. 
All kinematic measurements from VIRAC and \textit{Gaia} are in excellent agreement with the predictions from the fiducial barred model. Even though not fit to the VIRAC data, the model still explains 
\begin{inparaenum}
    \item all structures seen within the integrated maps,
    \item the RGB proper motion correlation in magnitude slices without the need for a separate classical bulge component, and
    \item the complex interplay of bar pattern rotation and streaming motions seen in the magnitude sliced mean proper motions.
\end{inparaenum}   
In future work we shall explore quantitatively the constraints on the pattern speed and mass distribution that can be obtained from VIRAC.
By fitting to the VIRAC data with the M2M method we shall obtain improved models for studying the detailed dynamics, and population dynamics, in the Galactic bulge.